\documentclass[11pt]{article}
\usepackage[utf8]{inputenc}
\pdfoutput = 1

\usepackage{amsmath}
\usepackage{amssymb}
\usepackage{bbm}
\usepackage{bbold}
\usepackage{braket}
\usepackage{caption}
\usepackage{cite}
\usepackage{color}
    \definecolor{darkgreen}{rgb}{0,0.5,0}
    \definecolor{darkred}{rgb}{0.5,0,0}
    \definecolor{darkblue}{rgb}{0,0,0.6}
    \definecolor{purple}{rgb}{0.4,.2,0.7}
    \definecolor{orange}{rgb}{0.7,0.3,0}
\usepackage{float}
\usepackage{graphicx}
\usepackage{fancyhdr}
\usepackage{tikz}
\usetikzlibrary{decorations.pathmorphing}
\usepackage{subcaption}
\usepackage[hyperfootnotes = true, colorlinks = true, linkcolor = blue, citecolor = red]{hyperref}
\usepackage{mathabx}
\usepackage[mathcal]{eucal}
\usepackage{pdfsync}
\usepackage{url}
\usepackage[normalem]{ulem}
\usepackage{comment}
\usepackage[english]{babel}
\usepackage{mathtools,slashed}
\usepackage{csquotes}
\usepackage{enumitem}
\usepackage[percent]{overpic}
\usepackage{epigraph} 
\usepackage{breqn}

\tikzset{snake it/.style={decorate, decoration=snake}}

\numberwithin{equation}{section}

\usepackage[margin = 2.2cm]{geometry}
\usepackage[ragged]{footmisc}
\renewcommand{\d}{\mathrm{d}}

\usepackage{color}
    \definecolor{darkgreen}{rgb}{0,0.5,0}
    \definecolor{darkred}{rgb}{0.5,0,0}
    \definecolor{darkblue}{rgb}{0,0,0.6}
    \definecolor{purple}{rgb}{0.4,.2,0.7}
    \definecolor{orange}{rgb}{0.7,0.3,0}
\usepackage{braket}
\usepackage{tikz}

\renewcommand{\d}{\mathrm{d}}

\newcommand{\sR}{\mathrm{R}}
\newcommand{\sL}{\mathrm{L}}
\newcommand{\RL}{\mathrm{R},\mathrm{L}}
\newcommand{\sC}{\mathrm{C}}
\newcommand{\sT}{\mathrm{T}}
\newcommand{\sM}{\mathrm{M}}
\newcommand{\sB}{\mathrm{B}}

\newcommand{\br}{\beta_{\sR}}
\newcommand{\bl}{\beta_{\sL}}

\newcommand{\btr}{\tilde{\beta}_{\sR}}
\newcommand{\btl}{\tilde{\beta}_{\sL}}
\newcommand{\cO}{\mathcal{O}}
\newcommand{\cOd}{\mathcal{O}^{\dagger}}
\newcommand{\cT}{\mathcal{T}}

\newcommand{\cW}{\mathcal{W}}

\newcommand{\dtau}{\Delta\tau}
\newcommand{\Veff}{V_{\mathrm{eff}}}
\newcommand{\Is}{I_{\mathrm{s}}}

\newcommand{\BH}{\mathrm{BH}}

\DeclareMathOperator{\Tr}{Tr}

\tikzset{cross/.style={cross out, draw=black, minimum size=2*(#1-\pgflinewidth), inner sep=0pt, outer sep=0pt},
cross/.default={1pt}}

\allowdisplaybreaks

\begin{document}

\thispagestyle{empty}
\begin{center}
    ~\vspace{5mm}
    
    {\LARGE \bf 
    Detecting Black Hole Microstates
    }
    
    \vspace{0.4in}
    
    {\bf Vijay Balasubramanian$^{1,2,3}$, William KL Chan$^1$, and Chitraang Murdia$^1$}

    \vspace{0.4in}

    $^1$ Department of Physics and Astronomy, University of Pennsylvania, Philadelphia, PA 19104, USA \vskip 1ex
    $^2$ Theoretische Natuurkunde, Vrije Universiteit Brussel (VUB) and The International Solvay Institutes, Pleinlaan 2, B-1050 Brussels, Belgium
    \vskip 1ex
    $^3$  Santa Fe Institute, 1399 Hyde Park Road, Santa Fe, NM 87501, USA
%    \vspace{0.1in}
    
    {\tt vijay@physics.upenn.edu, chanwill@sas.upenn.edu, murdia@sas.upenn.edu,}
\end{center}

\vspace{0.4in}

\begin{abstract}

We demonstrate that the Euclidean two-point function of an appropriately chosen probe operator can detect the microstate of an asymptotically AdS black hole.  This detection, which requires a tuned, state-dependent choice of probe, is the result of a new gravitational saddle, which dominates over the usual saddles.  The gravitational result can be explicitly reproduced in the dual boundary CFT if we assume the eigenstate thermalization hypothesis. We also discuss a  binary search protocol to detect the black hole microstate from a candidate list.

\end{abstract}

\tableofcontents

\section{Introduction}

Classically, there is no measurement at asymptotic infinity that detects the microstates that explain the Bekenstein-Hawking entropy of back holes. 
Indeed, this is why black hole spacetimes have classical horizons.  
Quantum mechanically we might hope that there would be some signal at infinity about the microstate, but any detection of such a signal must require non-perturbatively precise measurements or exponentially difficult computations \cite{Balasubramanian:2006iw,Brown:2019rox,Engelhardt:2021qjs,Balasubramanian:2022fiy, Engelhardt:2024hpe}.

In this work, we examine these issues in an explicit construction via the Euclidean path integral of black hole microstates, namely the shell microstates \cite{Balasubramanian:2022gmo,Balasubramanian:2022lnw,Climent:2024trz}.
These  microstates have coarse-grained descriptions as shells of dust propagating in the black hole interior, and are created in the dual field theory by the insertion of a heavy operator. 
The results in \cite{Balasubramanian:2025jeu,Balasubramanian:2025zey} show that these states span the complete Hilbert space of states of one- and two-boundary quantum gravity.  
Naively, they are overcomplete but non-perturbative effects in the gravitational path integral reduce the dimension of the space to precisely what we expect from the Bekenstein-Hawking formula.  
These results have been extended to include quantum corrections to the entropy \cite{He:2025neu} and out-of-equilibrium settings \cite{Balasubramanian:2024rek}. 
Thus the naive effective field theory (EFT) Hilbert space of the gravitational theory has enormously many null states, as expected, for example, from the toy model in \cite{Marolf:2020xie}. 
One consequence within the AdS/CFT correspondence is that the encoding of the bulk EFT of the black hole into the dual field theory will be non-isometric \cite{Akers:2022qdl}.  
Thus, reconstruction or detection of the microstate must be state-dependent \cite{Antonini:2024yif}.  
We will see this explicitly in our setting.

Specifically, we identify Euclidean computables that are sensitive to the shell microstates.
If we have a  microstate $\ket{\Psi_{\cO}}$ corresponding to the shell operator $\cO$, then the single-sided Euclidean two-point function $\left\langle \Psi_{\cO} \right| \mathbb{1}_L \otimes \psi(\tau_2) \psi^{\dagger}(\tau_1)_R \ket{\Psi_{\cO}}$ is sensitive to the choice of the probe operator $\psi$
For suitably chosen Euclidean times $\tau_1, \tau_2$, this two-point function has a huge spike if and only if the probe operator $\psi$ matches the shell operator $\cO$ , precisely as anticipated in \cite{Balasubramanian:2005kk}.
The key ingredient is  a new gravitational saddle -- dubbed the \emph{annihilation} saddle -- wherein two probe operators in the Euclidean past and future ``annihilate'' the corresponding shell operators.
This saddle is only present when the shell and probe operators are not orthogonal.
For appropriately tuned probe insertion times, the annihilation saddle dominates over the usual  \emph{propagation} saddle in which the shell and probe operators do not interact. The sharp peak in the Euclidean two-point function signals whether a black hole is in a given shell microstate $\ket{\Psi_{\cO}}$ or not. Given a finite list of candidate shell operators $\{ \cO_1, \dots, \cO_N \}$, we can then use a binary search to detect which operator corresponds to the black hole microstate.

In Section \ref{section:grav}, we use the gravitational path integral to compute the Euclidean two-point function and demonstrate the presence of the annihilation saddle for a suitable probe operator. We demonstrate this explicitly in two settings -- the limit of large shell mass in arbitrary dimensions and for any mass in the $(2+1)$-dimensional case.  We also discuss how to use a binary search protocol to detect the black hole microstate from a given set of candidates. In Section \ref{section:ETH}, we reach the same conclusions from the boundary perspective by assuming the Eigenstate Thermalization Hypothesis (ETH).
In Section \ref{section:noise}, we show that our results to robust to statistical variance arising from ensemble averaging.
We conclude by discussing some directions for the future in Section \ref{section:diss}.

\section{Gravity Calculation}
\label{section:grav}

We consider states in a theory of gravity with two asymptotic AdS boundaries with topology
$S^{d-1}\times \mathbb{R}$, where $\mathbb{R}$ represents time. Quantum gravity with
such boundary conditions is equivalent to the physics of two copies of a Conformal Field Theory (CFT), labeled $\mathrm{CFT}_{\sL}$ and $\mathrm{CFT}_{\sR}$, living at the left/right boundaries and with Hamiltonians $H_{\sL} = H_{\sR} = H$. We define the energy eigenbasis as
\begin{equation}
    H_L\ket{a,b} = E_a\ket{a,b}, \quad\quad H_R\ket{a,b} = E_b\ket{a,b}.
\end{equation} 

Following \cite{Balasubramanian:2022gmo,Balasubramanian:2022lnw}, a fixed-temperature, semiclassically well-controlled,  microstate of a two-sided black hole with independent inverse temperatures $\beta_{\sR}$ and $\beta_{\sL}$ can be constructed as
\begin{equation}
\label{eq:bh_microstate}
    \ket{\Psi_{\cO}} 
    = \ket{\rho_{\btl/2} \cO \rho_{\btr/2}}
    = \frac{1}{\sqrt{Z_1}} \sum_{a,b}e^{-\frac{1}{2}(\btl E_a + \btr E_b)}\cO_{ab}\ket{a,b} \, ,
\end{equation}
where the normalization is
\begin{equation} \label{eq:normalization}
    Z_1 = \Tr \left( \cOd e^{-\btl H}\cO e^{-\btr H} \right) \, .
\end{equation}
These fixed temperature states \eqref{eq:bh_microstate} are black hole microstates within the canonical ensemble.  The authors of  \cite{Balasubramanian:2025jeu,Balasubramanian:2025zey}  show  that these states form a basis, and also analyze the microcanonical and single-sided microstates.   See 
\cite{Barbon:2025bbh,Liu:2025xzd}
for discussion of the existence and stability of these and related states.
Note that the Euclidean times $\tilde{\beta}_{\RL}$ used to prepare the
states through Euclidean evolution are not necessarily equal to the physical inverse temperatures $\beta_{\mathrm{R,L}}$
of the black holes.

Let us describe the geometry associated with this state. 
We prepare the state in Euclidean signature and continue to Lorentzian signature at the $\tau=0$ slice.
On the Euclidean boundary, we insert the operator $\cO$ which creates a thin spherical dust shell. 
The mass of the shell is taken to be large,  
%(i.e., the corresponding operator is ``complex"), 
so the backreaction on the geometry is significant. 
This shell divides the Euclidean manifold into two connected components $X^{\RL}$ with geometries described by
\begin{equation}
    ds_{\RL}^2 = f_{\RL}(r)d\tau_{\RL}^2 + \frac{dr^2}{f_{\RL}(r)} + r^2 d\Omega_{d-1}^2,
\end{equation}
where 
\begin{equation}
\label{eq:metric-function}
    f_{\RL}(r) = 
    \begin{cases}
    r^2 - 8GM_{\RL} \, , & d = 2 \, , \\
    r^2 + 1 - \frac{16\pi GM_{\RL}}{(d-1)V_{\Omega}r^{d-2}} \, , & d > 2 \, . \\
    \end{cases}
\end{equation}
Here, $r$ is the Euclidean radial coordinate and $\tau_{\RL}$ are Euclidean times with the periodicity being the inverse temperature of the black hole $\beta_{\RL}$.\footnote{We work in units where the AdS radius $\ell=1$.} We define the black hole radii $r_{\sR}$ and $r_{\sL}$ as  solutions to the equations $f_{\sR}(r_{\sR}) = 0$ and $f_{\sL}(r_{\sL}) = 0$ respectively.

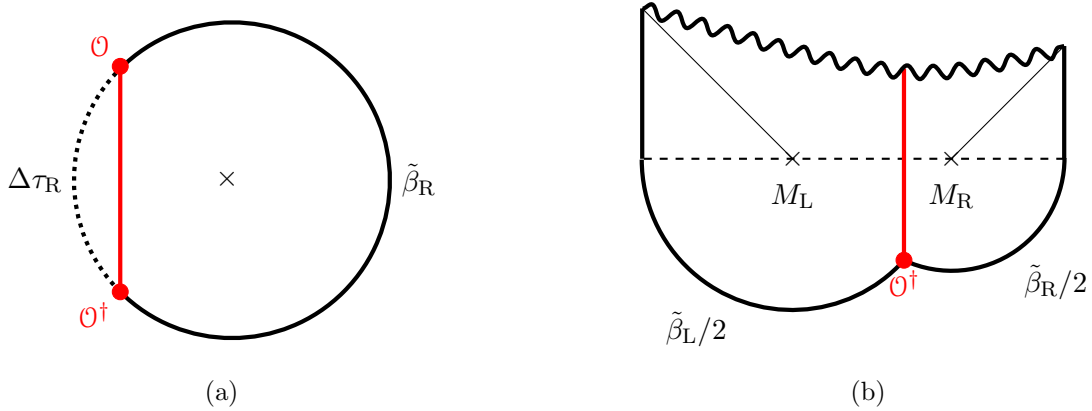
\begin{figure}
    \begin{subfigure}[t]{.5\textwidth}
        \centering
        \begin{tikzpicture}
            \draw[ultra thick] (-1.42,-1.5) arc (225:495:2.1) node[midway, right]{$\btr$};
            \draw[ultra thick, dotted] (-1.42,-1.5) arc (225:135:2.1) node[midway, left]{$\dtau_\sR$};
            \draw (0,0) node{$\times$};
            \draw[ultra thick, red] (-1.42,-1.5) -- (-1.42,1.5);
            \filldraw[red] (-1.42,1.5) circle (3pt) node[anchor=south east]{$\cO$};
            \filldraw[red] (-1.42,-1.5) circle (3pt) node[anchor=north east]{$\cOd$};
        \end{tikzpicture}            
        \caption{}
        \label{fig:shell-trajectory}
    \end{subfigure}%
    \hfill
    \begin{subfigure}[t]{.5\textwidth}
        \centering
        \begin{tikzpicture}
            \draw[ultra thick] (0,-1.35) arc (245:360:1.5) node[midway,anchor=north west]{$\btr/2$};
            \draw[ultra thick] (0,-1.35) arc (318:180:2) node[midway,anchor=north east]{$\btl/2$};
            \draw[thick,dashed] (2.13,0) -- (-3.48,0);
            \draw (-1.48,0) node{$\times$};
            \draw (0.63,0) node{$\times$};
            \draw (-1.48,-0.5) node{$M_\sL$};
            \draw (0.63,-0.5) node{$M_\sR$};
            \draw[ultra thick] (-3.48,0) -- (-3.48,2);
            \draw[ultra thick] (2.13,0) -- (2.13,1.5);
            \draw (-1.48,0) -- ((-3.48,2);
            \draw (0.63,0) -- (2.13,1.5);
            \filldraw[red] (0,-1.35) circle (3pt) node[anchor=north]{$\cOd$};
            \draw[ultra thick, red] (0,-1.35) -- (0,1.2);
            \path[draw=black, ultra thick, snake it] (-3.48,2) .. controls (0,1) .. (2.13,1.5);
        \end{tikzpicture}   
        \caption{}
        \label{fig:Lorentzian-continuation}
    \end{subfigure}%
    \caption{(a) The shell trajectory (red line) from the point of the operator insertion (red dot) on the right side of the Euclidean geometry, where we can see that the condition imposed by \eqref{eq:prep-time} is realized. (b) The state preparation beginning with the Euclidean geometry to prepare the two copies of the CFT up until the point of time-reflection symmetry, after which we continue to Lorentzian signature, resulting in the eternal Schwarzschild-AdS black hole geometry.}
    \label{fig:single-shell-state-preparation}
\end{figure}

The trajectory of the shell is given by $\left( r, \tau_{\RL} \right) = \left( R(T), \tau_{\RL}(T) \right)$ where $T$ is the shell's proper time which parametrizes this trajectory.
Since this shell serves as an interface between the left and right black holes, its trajectory should satisfy Israel's junction conditions \cite{Israel:1966rt},
\begin{equation} 
\label{eq:euc-eom}
    \left( \frac{dR}{dT} \right)^2 + \Veff(R) = 0.
\end{equation}
where
\begin{equation} 
\label{eq:eff-potential}
    \Veff(R) = -f_{\sR}(R) + \left( \frac{M_{\sR} - M_{\sL}}{m} - \frac{4\pi Gm}{(d-1)V_{\Omega}R^{d-2}} \right)^2,
\end{equation}
is the effective potential
and $V_{\Omega}$ is just the volume of the unit transverse sphere. 
The shell trajectory begins at the boundary as shown in Fig.~\ref{fig:shell-trajectory}, where $R=r_\infty$ and falls inwards toward the Euclidean horizon of the black hole. 
Upon reaching a minimum radius $R = R_*> r_{\sL}, r_{\sR}$, it bounces back toward the boundary at $r_\infty$, the bounce naturally being at the point of time-reflection symmetry. 
From the perspective of the right observer, the Euclidean time that passes throughout this journey of the shell is given by
\begin{equation} \label{eq:preparation-time}
    \dtau(\br;\bl) = 2\int_{R_*}^{r_\infty}\frac{dR}{f_{\sR}(R)}\sqrt{\frac{f_{\sR}(R) + \Veff(R)}{-\Veff(R)}}.
\end{equation}
Requiring that the backreaction of the shell gives the correct asymptotic mass on both sides, we get the relation between preparation times and inverse temperatures
\begin{equation}
    \label{eq:prep-time}
    \btl = \bl - \dtau(\bl;\br) \, , 
    \qquad    
    \btr = \br - \dtau(\br;\bl) \, .
\end{equation}
Note that this is precisely the condition we get from minimizing the action of the geometry.
This construction can easily be extended to the Lorentzian section, although our work in this article will only focus on  Euclidean signature. 

\begin{figure}[t]
    \centering
    \begin{tikzpicture}
        \draw[ultra thick] (0,1.5) arc (45:315:2.1) node[midway,left]{$\btl$};
        \draw[ultra thick] (0,-1.5) arc (225:495:2.1) node[midway,right]{$\btr$};
        \draw[ultra thick,red] (0,1.5) -- (0,-1.5);

        \filldraw[red] (0,1.5) circle (3pt) node[anchor=north east]{$\cO$};
        \filldraw[red] (0,-1.5) circle (3pt) node[anchor=south west]{$\cOd$};

        \draw (-1.8,0) node{$\bl$};
        \draw (1.8,0) node{$\br$};
    \end{tikzpicture}
    \caption{The gravitational saddle corresponding to $Z_1$ in \eqref{eq:normalization}. This depicts the Euclidean disk geometries of the left and right black holes being glued along the shell trajectory.}
    \label{fig:grav-saddle}
\end{figure}
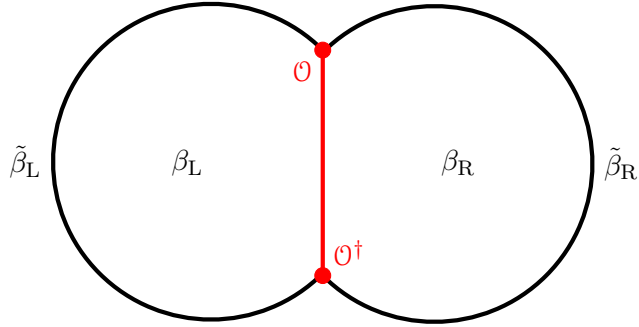

The normalization $Z_1$ can be computed using the action of the gravitational saddle, which is sketched in Fig.~\ref{fig:grav-saddle}.
The Euclidean action corresponding to this saddle is a sum of three terms -- the action of the two disks and the action associated with the shell
\begin{equation} \label{eq:action}
    I_E = \btl F(\bl) +  \btr F(\br) + \Is(\br,\bl) \, .
\end{equation}
Here, $F(\beta)$ is the renormalized free energy of the black hole given by \cite{Emparan:1999pm}
\begin{equation} 
\label{eq:free-energy}
    F(\beta) 
    = - \frac{1}{\beta}\log Z(\beta) 
    = \frac{V_{\Omega}}{16\pi G}(-r_{\beta}^d + r_{\beta}^{d-2} + c_d) \, ,
\end{equation}
where $r_\beta$ is the black hole radius at inverse temperature $\beta$ and $c_d$ accounts for the Casimir energy of the CFT in even dimensions \cite{Balasubramanian:1999re}.
Also, $\Is(\br,\bl)$ is the renormalized action associated with the shell
\begin{equation}
\label{eq:shell-action}
    \Is(\br,\bl) 
    = \frac{d}{8\pi G} \text{Vol}_{s,\sR} + \frac{d}{8\pi G} \text{Vol}_{s,\sL} + m\frac{d-2}{d-1}L[\gamma_\cW] \, .
\end{equation}
Here we have expressed the action of the shell as the sum of three terms -- the volume as seen on the right side, the volume as seen on the left side, and the term containing the proper length of the shell's trajectory $L[\gamma_\cW]$.
Explicitly, these expressions are
\begin{align}
    \text{Vol}_{s,\sR} 
    &= \frac{2V_\Omega}{d} \int_{R_{*}}^{r_{\infty}} \frac{dR}{f_{\sR}(R)}\sqrt{\frac{f_{\sR}(R)+\Veff(R)}{-\Veff(R)}}(R^d - r_{\sR}^d) \, , \\
    \text{Vol}_{s,\sL} 
    &= \frac{2V_\Omega}{d} \int_{R_{*}}^{r_{\infty}} \frac{dR}{f_{\sL}(R)}\sqrt{\frac{f_{\sL}(R)+\Veff(R)}{-\Veff(R)}}(R^d - r_{\sL}^d) \, , \\
    L[\gamma_\cW] 
    &= 2\int_{R_{*}}^{r_{\infty}}\frac{dR}{\sqrt{-\Veff(R)}} \, ,
\end{align}
where we will take the limit $r_{\infty} \to \infty$.
We obtain the normalization
\begin{equation}
\label{eq:norm-partition-fn}
    Z_1 
    = e^{-I_E} 
    = \exp \left( -  \btl F(\bl) - \btr F(\br) - \Is(\br,\bl) \right) \, .
\end{equation}

\subsection{Detecting microstates}

The states constructed above have bulk duals that contribute to the semi-classical black hole Hilbert space and thus are valid microstates.  Indeed, they form a complete basis, as shown in \cite{Balasubramanian:2025jeu,Balasubramanian:2025zey, Balasubramanian:2025hns}. A natural question to ask is whether or not these microstates can be detected by asymptotic correlation functions.  The reason to think this might be possibe is that, while the classical geometry outisde the horizon contains no trace of the shell behind it, the quantum state does \cite{Balasubramanian:2022gmo}, so that there is a form of ``quantum hair''. In this section, we address this question by computing the single-sided Euclidean two-point function\footnote{A related computation that is not sensitive to the black hole microstate is the thermal  two-point function
$\frac{1}{\Tr e^{-\btr H}} \Tr \left( e^{-\btr H} \psi(\tau_2) \psi^{\dagger}(\tau_1) \right)$. }
\begin{equation} 
\label{eq:two-pt-function}
    \left\langle \Psi_{\cO} \right| \mathbb{1}_L \otimes \psi(\tau_2) \psi^{\dagger}(\tau_1)_R \ket{\Psi_{\cO}} 
    = \frac{1}{Z_1} \Tr \left( \cOd e^{-(\btr/2) H} \psi(\tau_2) \psi^{\dagger}(\tau_1) e^{-(\btr/2) H} \cO e^{-\btl H} \right) \, , 
\end{equation}
and showing how this quantity depends on the relationship between the interior shell operator $\cO$ and the probe operator $\psi$. 

Consider an arbitrary  probe of the form
\begin{equation} \label{eq:op-complex}
    \psi = w_{\cO} \cO + \sum_{i} w_i \phi_i \, ,
\end{equation}
where $w_{*} \in [0,1]$ represent weights for the operators and $\{\cO, \phi_i \}$ is some orthonormal basis of operators.  (We have absorbed phases in coefficients into the definitions of the operators.)
One way to construct such a basis is to begin with a complete set of shell states, and to then use a Gram-Schmidt procedure to obtain an orthogonal set of states. We can insist that ${\cal O}$ is a specific shell operator if we want, but  many of the $\phi_i$ operators will no longer be shell operators, but will rather be complicated linear combinations constructed from them.
However, the details of these $\phi_i$ operators are not important to us. All we require is that an orthonormal basis exists with $\cO$ being a basis element, so any operator $\psi$ can be decomposed into a component along $\cO$ and a component in its orthogonal complement. 

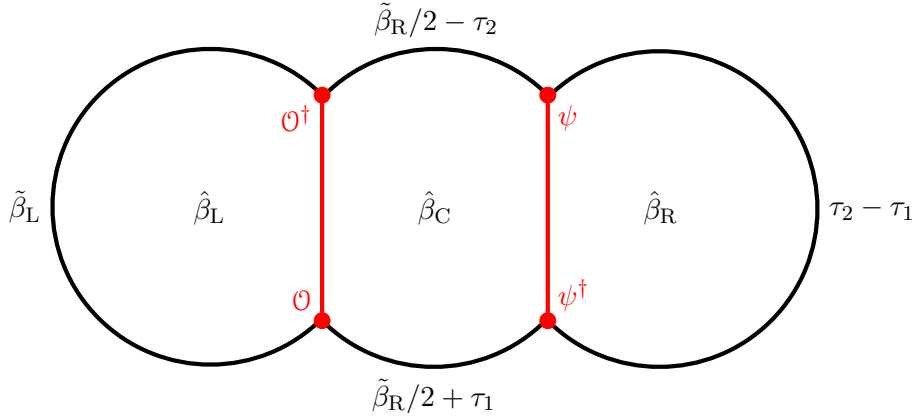
\begin{figure}[t]
    \centering
    \begin{tikzpicture}
        \draw[ultra thick] (0,1.5) arc (45:315:2.1) node[midway,left]{$\btl$};
        \draw[ultra thick] (3,1.5) arc (45:135:2.1) node[midway,above]{$\btr/2 - \tau_2$};
        \draw[ultra thick] (3,-1.5) arc (225:495:2.1) node[midway,right]{$\tau_2 - \tau_1$};
        \draw[ultra thick] (0,-1.5) arc (225:315:2.1) node[midway,below]{$\btr/2 + \tau_1$};
        \filldraw[red] (0,1.5) circle (3pt) node[anchor=north east]{$\cO^{\dagger}$};
        \filldraw[red] (0,-1.5) circle (3pt) node[anchor=south east]{$\cO$};
        \filldraw[red] (3,1.5) circle (3pt) node[anchor=north west]{$\psi$};
        \filldraw[red] (3,-1.5) circle (3pt) node[anchor=south west]{$\psi^{\dagger}$};
        \draw[ultra thick, red] (0,1.5) -- (0,-1.5);
        \draw[ultra thick, red] (3,1.5) -- (3,-1.5);
        \draw (1.5,0) node{$\hat{\beta}_{\sC}$};
        \draw (-1.5,0) node{$\hat{\beta}_{\sL}$};
        \draw (4.5,0) node{$\hat{\beta}_{\sR}$};
    \end{tikzpicture}
    \caption{A sketch of the propagation saddle.
    In this saddle, the interior shell and probe do not interact with each other, so this saddle exists for any choice of the probe operator.}
    \label{fig:propagation-saddle}
\end{figure}

\begin{figure}[t]
    \centering
    \begin{tikzpicture}
        \draw[ultra thick] (0,1.5) arc (100:260:1.55) node[midway,left]{$\btl$};
        \draw[ultra thick] (1.5,1.5) arc (315:585:1.1) node[midway,above]{$\btr/2 - \tau_2$};
        \draw[ultra thick] (1.5,-1.5) arc (280:440:1.55) node[midway,right]{$\tau_2 - \tau_1$};
        \draw[ultra thick] (0,-1.5) arc (135:405:1.1) node[midway,below]{$\btr/2 + \tau_1$};
        \filldraw[red] (0,1.5) circle (3pt) node[anchor=north east]{$\cO^{\dagger}$};
        \filldraw[red] (0,-1.5) circle (3pt) node[anchor=south east]{$\cO$};
        \filldraw[red] (1.5,1.5) circle (3pt) node[anchor=north west]{$\cO$};
        \filldraw[red] (1.5,-1.5) circle (3pt) node[anchor=south west]{$\cO^{\dagger}$};
        \draw[ultra thick, red] (0,1.5) .. controls (0.5,1.4) and (1,1.4) .. (1.5,1.5);
        \draw[ultra thick, red] (0,-1.5) .. controls (0.5,-1.4) and (1,-1.4) .. (1.5,-1.5);
        \draw (0.7,0) node{$\hat{\beta}_{\sM}$};
        \draw (0.75,2.5) node{$\hat{\beta}_{\sT}$};
        \draw (0.75,-2.5) node{$\hat{\beta}_{\sB}$};
    \end{tikzpicture}
    \caption{A sketch of the annihilation saddle. In this saddle, the two probe operators in the Euclidean past and future ``annihilate'' the corresponding interior shell operators. This saddle only exists if the probe operator matches the interior shell operator, so it is crucial for detectability.}
    \label{fig:annihilation-saddle}
\end{figure}
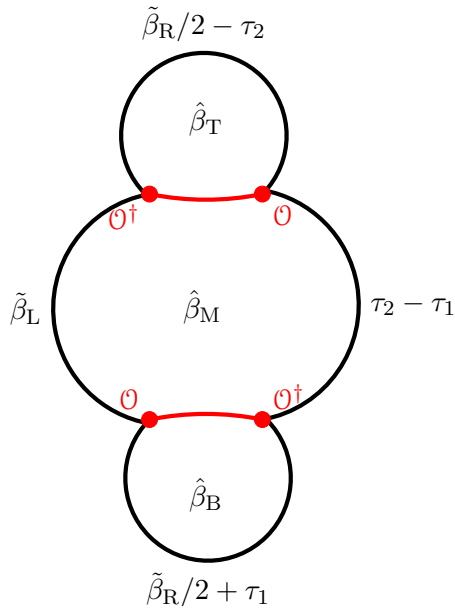

We can assume that the probe operator is  normalized, so the weights satisfy  $w_{\cO}^2 + \sum_{i} w_i^2 = 1$.
It follows that the numerator of the two-point function in \eqref{eq:two-pt-function} can be expressed as
\begin{multline} 
\label{eq:t1-complex-op}
    \mathcal{T}_1 = w_{\cO}^2 \Tr \left( \cOd e^{-\btr H/2 } \cO(\tau_2) \cO^{\dagger}(\tau_1) e^{-\btr H/2} \cO e^{-\btl H} \right) \\
    + \sum_{i} w_i^2 \Tr \left( \cOd e^{-\btr H/2} \phi_i(\tau_2) \phi_i^{\dagger}(\tau_1) e^{-\btr H/2} \cO e^{-\btl H} \right) \, .
\end{multline}
We can evaluate these traces using gravitational saddle point analysis.
For the second term in \eqref{eq:t1-complex-op}, the gravitational saddle is shown in Fig.~\ref{fig:propagation-saddle} --  we call this the \emph{propagation} saddle.
For the first term in \eqref{eq:t1-complex-op}, there are two possible saddles - the propagation saddle, and another one shown in Fig.~\ref{fig:annihilation-saddle}, which we call the \emph{annihilation} saddle.
The annihilation saddle allows us to detect the state, since it exists if and only if the probe operator matches the interior shell operator. 

The Euclidean actions for each of these saddles can be computed as in \eqref{eq:action}.
The actions for the propagation and the annihilation saddles respectively, are
\begin{align}
\label{eq:propagation-saddle-action}
    I_P
    &= \tilde{\beta}_{\sL} F(\hat{\beta}_{\sL}) + \tilde{\beta}_{\sC} F(\hat{\beta}_{\sC}) + \tilde{\beta}_{\sR} F(\hat{\beta}_{\sR}) + I_{s}(\hat{\beta}_{\sL}, \hat{\beta}_{\sC}) + I_{s}(\hat{\beta}_{C}, \hat{\beta}_{\sR}) \, , \\
\label{eq:annihilation-saddle-action}
    I_A 
    &= \tilde{\beta}_{\sT} F(\hat{\beta}_{\sT}) + \tilde{\beta}_{\sM} F(\hat{\beta}_{\sM}) + \tilde{\beta}_{\sB} F(\hat{\beta}_{\sB}) + I_{s}(\hat{\beta}_{\sT}, \hat{\beta}_{\sM}) + I_{s}(\hat{\beta}_{\sM}, \hat{\beta}_{\sB}) \ \, .
\end{align}
where the physical inverse temperatures $\hat{\beta}_i$'s appearing in these equations must be computed by solving the corresponding saddle point equations.
The equations for the propagation saddle are
\begin{equation}
\begin{split}
\label{eq:grav-propagation-saddle}
    \hat{\beta}_{\sL} &= \btl + \dtau(\hat{\beta}_{\sL}; \hat{\beta}_{\sC}) \, , \\
    \hat{\beta}_{\sC} &= \btr - (\tau_2 - \tau_1) + \dtau(\hat{\beta}_{\sC}; \hat{\beta}_{\sL}) + \dtau(\hat{\beta}_{\sC}; \hat{\beta}_{\sR}) \, , \\
    \hat{\beta}_{\sR} &= \tau_2 - \tau_1 + \dtau(\hat{\beta}_{\sR}; \hat{\beta}_{\sC}) \, ,
\end{split}
\end{equation}
and for the annihilation saddle are 
\begin{equation}
\begin{split}
\label{eq:grav-annihilation-saddle}
    \hat{\beta}_{\sT} &= \btr/2 - \tau_2 + \dtau(\hat{\beta}_{\sT}; \hat{\beta}_{\sM}) \, , \\
    \hat{\beta}_{\sM} &= \btl + \tau_2 - \tau_1 + \dtau(\hat{\beta}_{\sM}; \hat{\beta}_{\sT}) + \dtau(\hat{\beta}_{\sM}; \hat{\beta}_{\sB}) \, , \\
    \hat{\beta}_{\sB} &= \btr/2 + \tau_1 + \dtau(\hat{\beta}_{\sB}; \hat{\beta}_{\sM}) \, .
\end{split}
\end{equation}

Thus, we get the final result
\begin{equation}
    \cT_1 = \cT_1^{P} + w_{\cO}^2 \cT_1^{A}
\end{equation}
where $\cT_1^{P} \sim e^{-I_P}$ and $\cT_1^{A} \sim e^{-I_A}$ are the contribution from the propagation and annihilation saddles respectively. 

To detect the interior state, the annihilation saddle should dominate over the propagation saddle. 
This can be achieved by tuning $\tau_1$ and $\tau_2$.
If $\btr/2 + \tau_1$ or $\btr/2 - \tau_2$ is made sufficiently small, the respective free energies, $F(\hat{\beta}_{\sT})$ or $F(\hat{\beta}_{\sB})$, become large. 
By making either one or both of these contributions large, we can always tune $\cT_1^{A}$ to be significantly larger than $\cT_1^{P}$.  Below, we demonstrate explicitly in two representative cases that the Euclidean correlators have a large enough signal to detect the interior state. It is worth mentioning that the Lorentzian correlators also receive a small contribution from the annihilation saddle, but this signal is small and noisy; see \cite{Balasubramanian:2025akx} for an analysis of the Lorentzian setting. 

\begin{figure}[t]
    \centering
    \begin{tikzpicture}
        \draw[ultra thick] (0,1.5) arc (45:315:2.1) node[midway,left]{$\btl$};
        \draw[ultra thick] (0,-1.5) arc (225:495:2.1) node[midway,right]{$\btr$};
        \draw[ultra thick,red] (0,1.5) -- (0,-1.5);

        \filldraw[red] (0,1.5) circle (3pt) node[anchor=north east]{$\cO$};
        \filldraw[red] (0,-1.5) circle (3pt) node[anchor=south west]{$\cOd$};

        \draw[thick,red,dotted] (3,1.5) .. controls (2.5,0) .. (3,-1.5);

        \filldraw[red] (3,1.5) circle (3pt) node[anchor=south west]{$\psi^{\dagger}_{\text{light}}$};
        \filldraw[red] (3,-1.5) circle (3pt) node[anchor=north west]{$\psi_{\text{light}}$};

        \draw (-1.8,0) node{$\bl$};
        \draw (1.8,0) node{$\br$};
    \end{tikzpicture}
    \caption{Insertion of a light operator which does not backreact on the geometry cannot detect the black hole microstate. The resulting gravitational saddle is identical to Fig.~\ref{fig:grav-saddle}.}
    \label{fig:light-operator}
\end{figure}
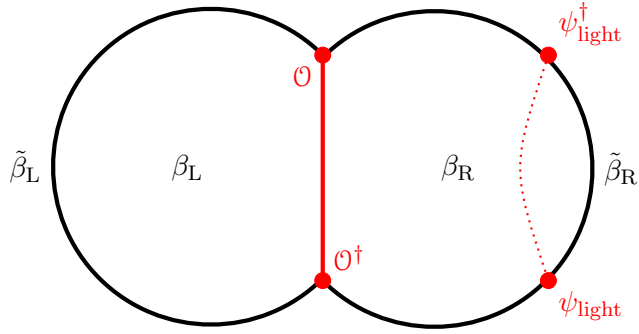

Finally, consider also the case where the probe operator $\psi$ is light, so that it does not interact with the shell operator $\cO$ in an operator-dependent way.
Because the probe is light, there is negligible backreaction on the geometry and so the operator never sees the black hole microstate - this is depicted in Fig.~\ref{fig:light-operator}.
Detection of the black hole microstate depends entirely  on the existence of the annihilation saddle, which is absent now as the first term in \eqref{eq:t1-complex-op} does not arise. 
Hence, we conclude that a light operator cannot detect the black hole microstate.

\subsection{Large mass limit}

We first demonstrate that one can detect a particular black hole microstate explicitly in the  limit of large shell rest mass,  $m \rightarrow \infty$.
This is the limit in which the shell mass significantly exceeds the asymptotic mass of the black hole.
Crucially, in this limit, the shell action \eqref{eq:shell-action} no longer depends on the inverse temperatures of the Euclidean geometries \cite{Balasubramanian:2022gmo},
\begin{equation}
\label{eq:shell-action-large-mass}    
    \Is(\beta_R,\beta_L) \approx \Is(m) = 2 m \log R_* \, .
\end{equation}
where $R_* \sim G m$.
Moreover, the Euclidean time elapsed by the shell trajectory vanishes, $\dtau(\beta_R; \beta_L) \approx 0$.
Therefore, using \eqref{eq:prep-time}, it follows that $\btl\approx\bl$ and $\btr\approx\br$ respectively.

We can now evaluate the saddle point contributions to the two-point function. 
Each contribution can be expressed as a product of partition functions for the disks and shells involved in each diagram. This simplification arises because the shell action $\Is$ is independent of the disk parameters. 
Therefore, the normalization factor simplifies to
\begin{equation}
    Z_1 \approx Z( \tilde{\beta}_{\sL}) Z( \tilde{\beta}_{\sR}) e^{- \Is(m)} \, ,
\end{equation}
where we have omitted  subleading contributions, including the one-loop corrections.
For the annihilation and propagation saddles, the saddle point equations are
\begin{align}
    \label{eq:a-saddle-large-mass}
    \hat{\beta}_{\sT} = \btr/2 - \tau_2 \, ,
    \qquad
    \hat{\beta}_{\sM} = \btl + \tau_2 - \tau_1 \, ,
    \qquad
    \hat{\beta}_{\sB} = \btr/2 + \tau_1 \, . \\
    \label{eq:p-saddle-large-mass}
    \hat{\beta}_{\sC} = \btr - (\tau_2 - \tau_1) \, ,
    \qquad \qquad
    \hat{\beta}_{\sR} = \tau_2 - \tau_1 \, , 
    \qquad \qquad
    \hat{\beta}_{\sL} = \btl \, .
\end{align}
Thus, these saddle point contributions simplify to 
\begin{align}
    \label{eq:saddle-disk-propagation}
    \cT^{P}_1 
    &\approx  Z(\hat{\beta}_{\sL}) Z(\hat{\beta}_{\sC}) Z(\hat{\beta}_{\sR}) e^{- 2 \Is(m)} \, , \\
    \label{eq:saddle-disk-annihilation}
    \cT^{A}_1 
    &\approx Z(\hat{\beta}_{\sT}) Z(\hat{\beta}_{\sM}) Z(\hat{\beta}_{\sB}) e^{- 2 \Is(m)} \, .
\end{align}
The explicit answers for $(d+1)$-dimensions are
\begin{align} 
\begin{split} 
\label{eq:grav-propagation-evaluation}
    \log \cT^{P}_1 \approx \frac{V_\Omega}{8 G} \Biggl[ & \left( \frac{2 \pi}{\btl} \right)^{d-1} - \left( \frac{2 \pi}{\btl} \right)^{d-3} + \left( \frac{2 \pi}{\btr + \tau_1 - \tau_2} \right)^{d-1} - \left( \frac{2 \pi}{\btr + \tau_1 - \tau_2} \right)^{d-3} \\ 
    &\hspace{0.8in} + \left( \frac{2 \pi}{\tau_2 - \tau_1} \right)^{d-1} - \left( \frac{2 \pi}{\tau_2 - \tau_1} \right)^{d-3} + 2\pi c_d (\btl + \btr) \Biggr] -  2 \Is(m) \, , 
\end{split} \\
\begin{split} 
\label{eq:grav-annihilation-evaluation}
    \log \cT^{A}_1 \approx \frac{V_\Omega}{8 G} \Biggl[ &  \left( \frac{2 \pi}{\frac{\btr}{2} - \tau_2} \right)^{d-1} - \left( \frac{2 \pi}{\frac{\btr}{2} - \tau_2} \right)^{d-3} + \left( \frac{2 \pi}{\btl + \tau_2 - \tau_1} \right)^{d-1} - \left( \frac{2 \pi}{\btl + \tau_2 - \tau_1} \right)^{d-3}   \\ 
    & \hspace{0.8in} + \left( \frac{2 \pi}{\frac{\btr}{2} + \tau_1} \right)^{d-1} - \left( \frac{2 \pi}{\frac{\btr}{2} + \tau_1} \right)^{d-3}+ 2\pi c_d (\btl + \btr) \Biggr] - 2 \Is(m) \, ,
\end{split} 
\end{align}
where the first term on the right side comes from the disk actions and the second term comes from the shell actions.
For detection, $\mathcal{T}_{1}^{A}$ should be significantly larger than $\mathcal{T}_{1}^{P}$.
This can be achieved by tuning the Euclidean insertion times $\tau_1$ and $\tau_2$ for the probe operators.
Note that the annihilation saddle contribution $\mathcal{T}_{1}^{A}$ has a divergence at $\tau_2 = -\tau_1 = \frac{\btr}{2}$, so we can make it sufficiently larger than $\mathcal{T}_{1}^{P}$ by choosing
$\tau_1$ and $\tau_2$ close to this divergence.
Apart from the Hawking temperature, no specific knowledge about the shell operator is needed for this tuning.
As mentioned, this tuning does not work in the Lorentzian correlators -- the divergence in question can only be achieved in the Euclidean correlator. 

More concretely, the explicit answers in the simplest case of $(2+1)$-dimensions are
\begin{align}
\label{eq:propagation-contribution}
    \log \cT^{P}_1 &\approx \frac{\pi^2}{4G}\left( \frac{1}{\btr - (\tau_2 - \tau_1)} + \frac{1}{\tau_2 - \tau_1} + \frac{1}{\btl} \right) + \frac{1}{32G}(\btl + \btr) + 4m\log G m \, , \\
\label{eq:annihilation-contribution}
    \log \cT^{A}_1 &\approx \frac{\pi^2}{4G}\left( \frac{1}{\btr/2 - \tau_2} + \frac{1}{\btr/2 + \tau_1} + \frac{1}{\btl + \tau_2 - \tau_1} \right) + \frac{1}{32G}(\btl + \btr) + 4m\log G m \, .
\end{align}
The switchover of dominance between these saddles occurs when the difference,
\begin{multline}
\label{eq:saddle-diff}
    \log \cT^{P}_1 - \log \cT^{A}_1 \approx \frac{\pi^2}{4G} \bigg[ \left( \frac{1}{\btr - (\tau_2 - \tau_1)} + \frac{1}{\tau_2 - \tau_1} + \frac{1}{\btl} \right) \\
    - \left( \frac{1}{\btr/2 - \tau_2} + \frac{1}{\btr/2 + \tau_1} + \frac{1}{\btl + \tau_2 - \tau_1} \right) \bigg] \, ,
\end{multline}
vanishes.  When this difference is positive, the propagation saddle dominates; otherwise, the annihilation saddle dominates.
Clearly, rescaling all parameters $\btr$, $\btl$, $\tau_1$, $\tau_2$ by the same factor does not affect the sign of this difference. 
Thus, we can set $\btr = 1$ and then study the effects of varying the other parameters.

\begin{figure}
    \begin{subfigure}[b]{.24\textwidth}
        \centering
        \includegraphics[scale=.165]{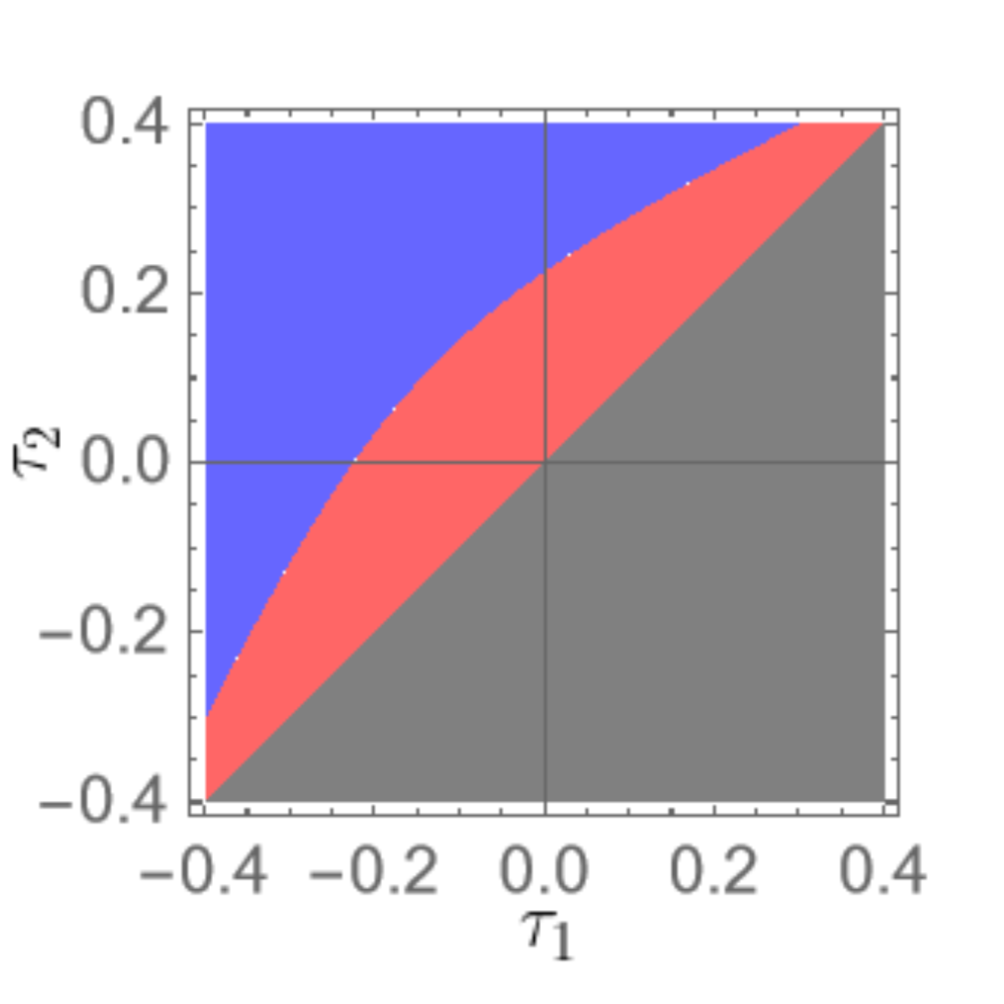}
        \caption{$\btl = 10$}
        \label{fig:btl-large}
    \end{subfigure}%
    \hfill
    \begin{subfigure}[b]{.24\textwidth}
        \centering
        \includegraphics[scale=.165]{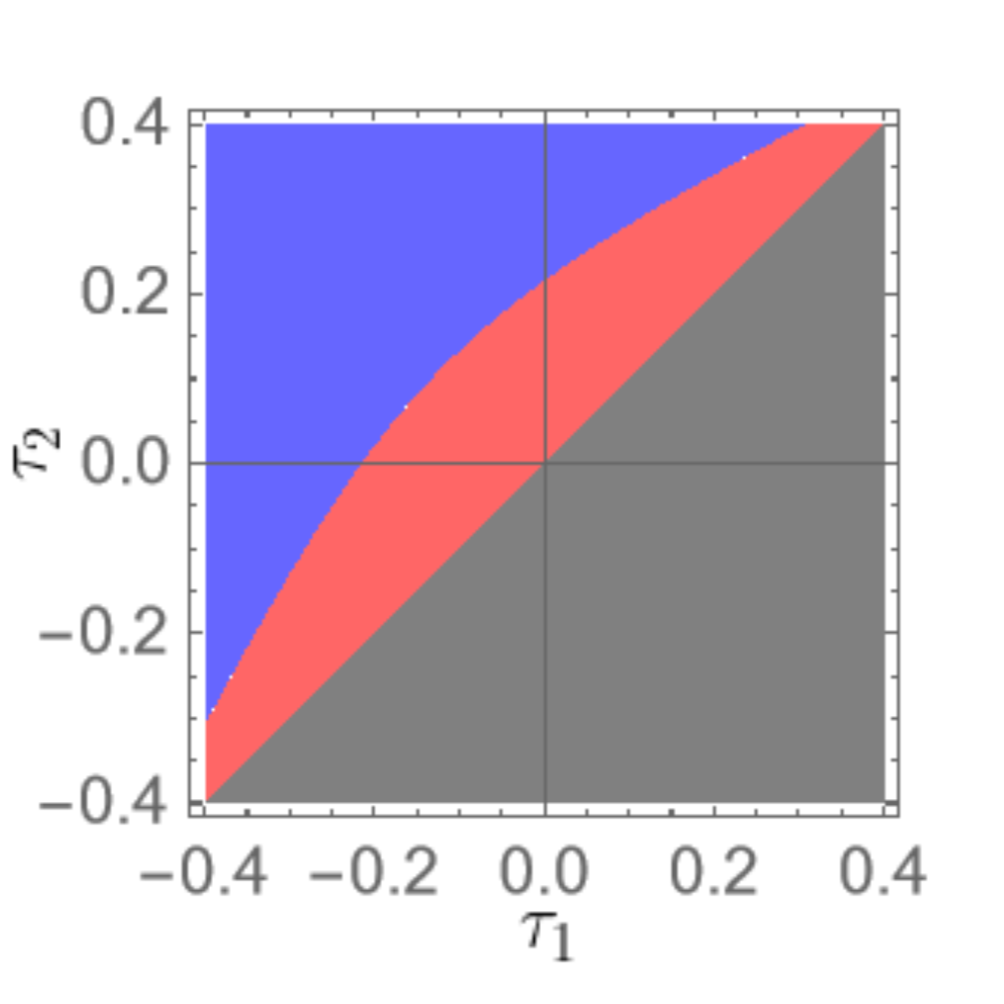}
        \caption{$\btl = 1$}
        \label{fig:btl=btr}
    \end{subfigure}%
    \hfill 
    \begin{subfigure}[b]{.24\textwidth}
        \centering
        \includegraphics[scale=.16]{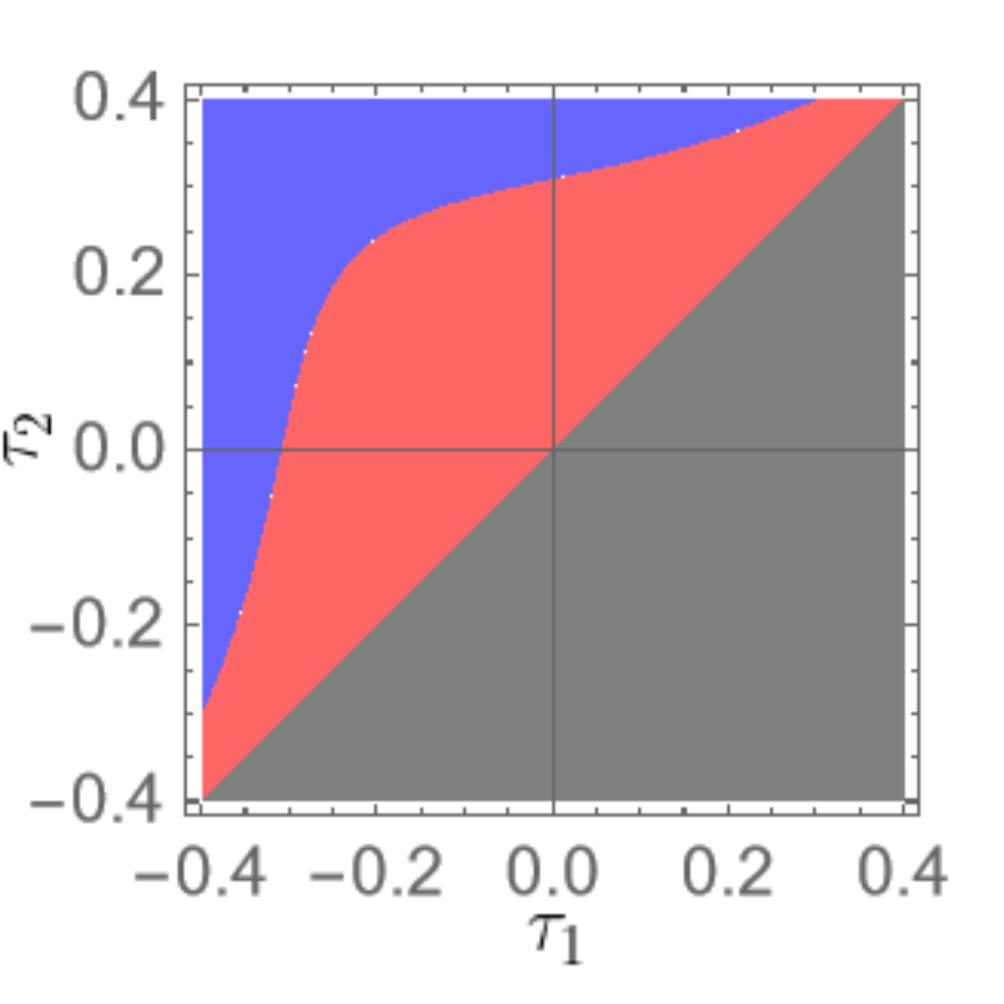}
        \caption{$\btl = 0.2$}
        \label{fig:btr-large}
    \end{subfigure}%
    \hfill
    \begin{subfigure}[b]{.24\textwidth}
        \centering
        \includegraphics[scale=.16]{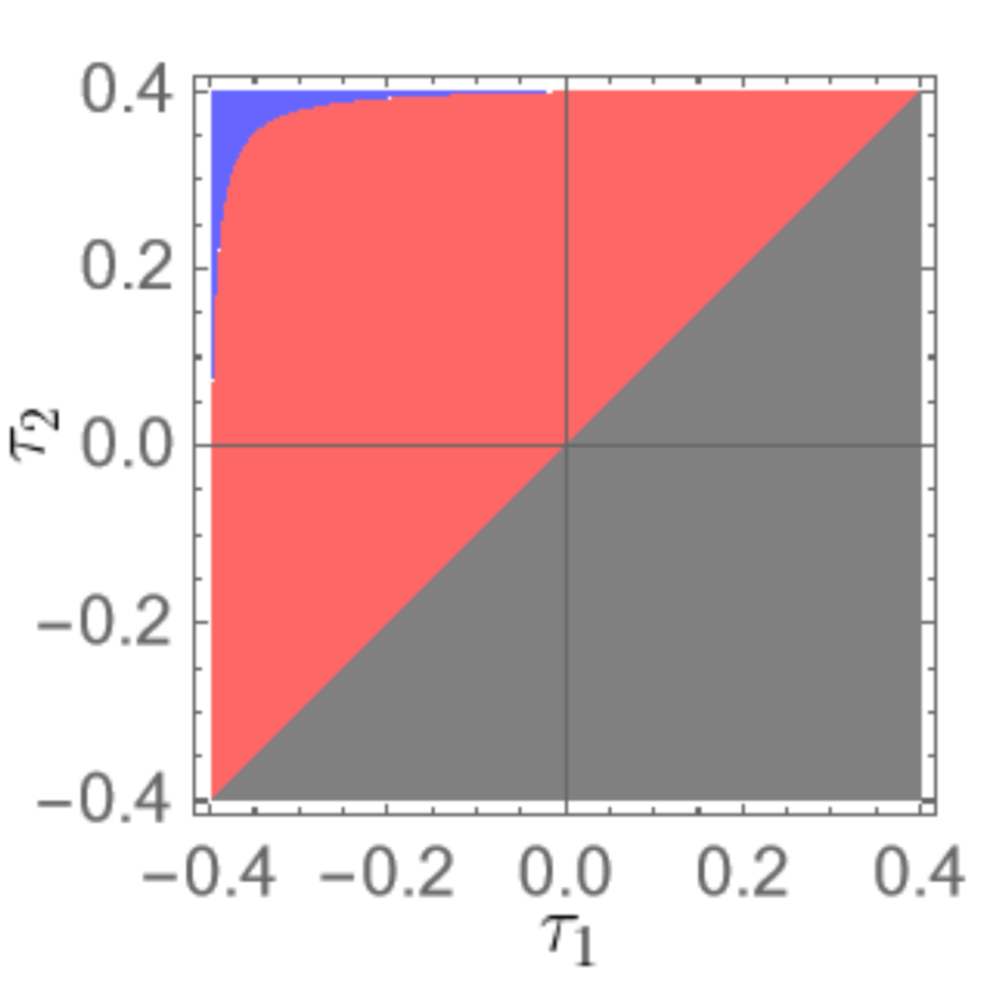}
        \caption{$\btl = 0.1$}
        \label{fig:btr-largest}
    \end{subfigure}
    \caption{The switchover of dominance between the annihilation and propagation saddle in the large mass limit $m\rightarrow\infty$. We set $\btr = 1$ and vary $\btl$ to see which saddle dominates as a function of the Euclidean preparation times $\tau_1$, $\tau_2$. The annihilation saddle dominates in the blue region, whereas the propagation saddle dominates in the red region. The gray area with $\tau_2 <  \tau_1$ is forbidden.}
    \label{fig:dominance-4plot}
\end{figure}

Fig.~\ref{fig:dominance-4plot} shows the dominant saddle as a function of $\tau_1$ and $\tau_2$ at various values of $\btl$.
When $\btl$ is large, i.e. $\btl \approx (\btl + \tau_2 - \tau_1)$, the $\btl$-dependent contributions to the propagation and annihilation saddle are almost equal.
Thus, $\btl$ drops out of the difference in \eqref{eq:saddle-diff} and the switchover of dominance is not sensitive  to it. 
This feature can be seen in Figs.~\ref{fig:btl-large}, \ref{fig:btl=btr} which are almost identical. 
Physically, this makes sense because the left Euclidean boundary is far from the interior shell, so it does not affect the interaction between the shell and probe operators. 
Now consider the other scenario where we make $\btl$ small.
In this limit, the propagation saddle dominates because interior shell operators prefer to connect to each other rather than to the probe operators as needed for the annihilation saddle.
Thus, it gets harder to detect the microstate and we need to choose larger values for the Euclidean preparation times $(-\tau_1)$, $\tau_2$ for the annihilation saddle to dominate.
This can be seen in Figs.~\ref{fig:btl=btr}, \ref{fig:btr-large},  \ref{fig:btr-largest}, where the annihilation saddle dominates over a smaller region in parameter space as we increase $\btl$. 

\subsection{(2+1)-dimensional case for generic mass}

\begin{figure}
    \begin{subfigure}[b]{.33\textwidth}
        \centering
        \includegraphics[scale=.2]{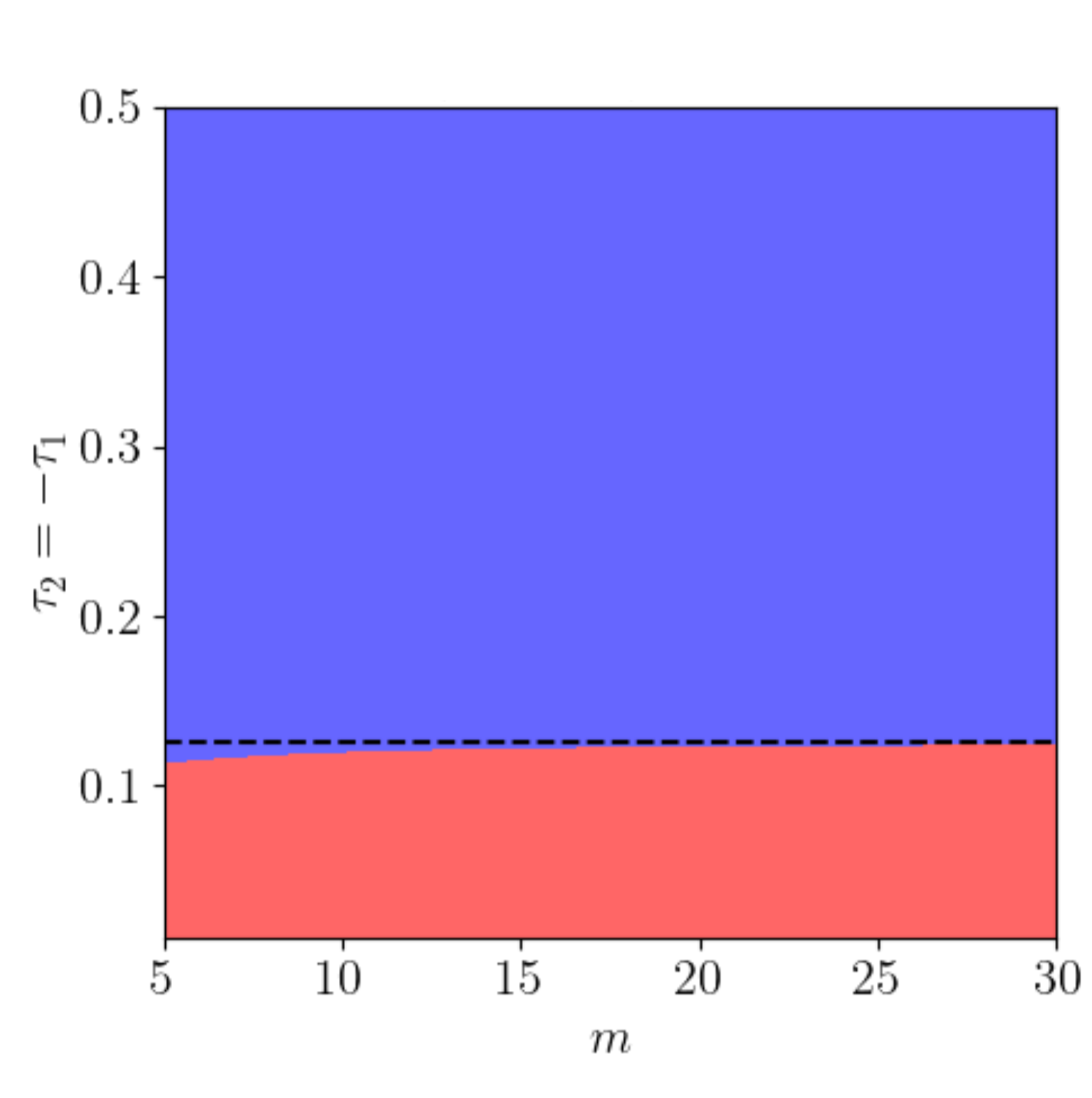}
        \caption{$\btr = 1$, $\btl = 5$}
        \label{fig:temp-ratio-one-fifth}
    \end{subfigure}%
    \hfill 
    \begin{subfigure}[b]{.33\textwidth}
        \centering
        \includegraphics[scale=.2]{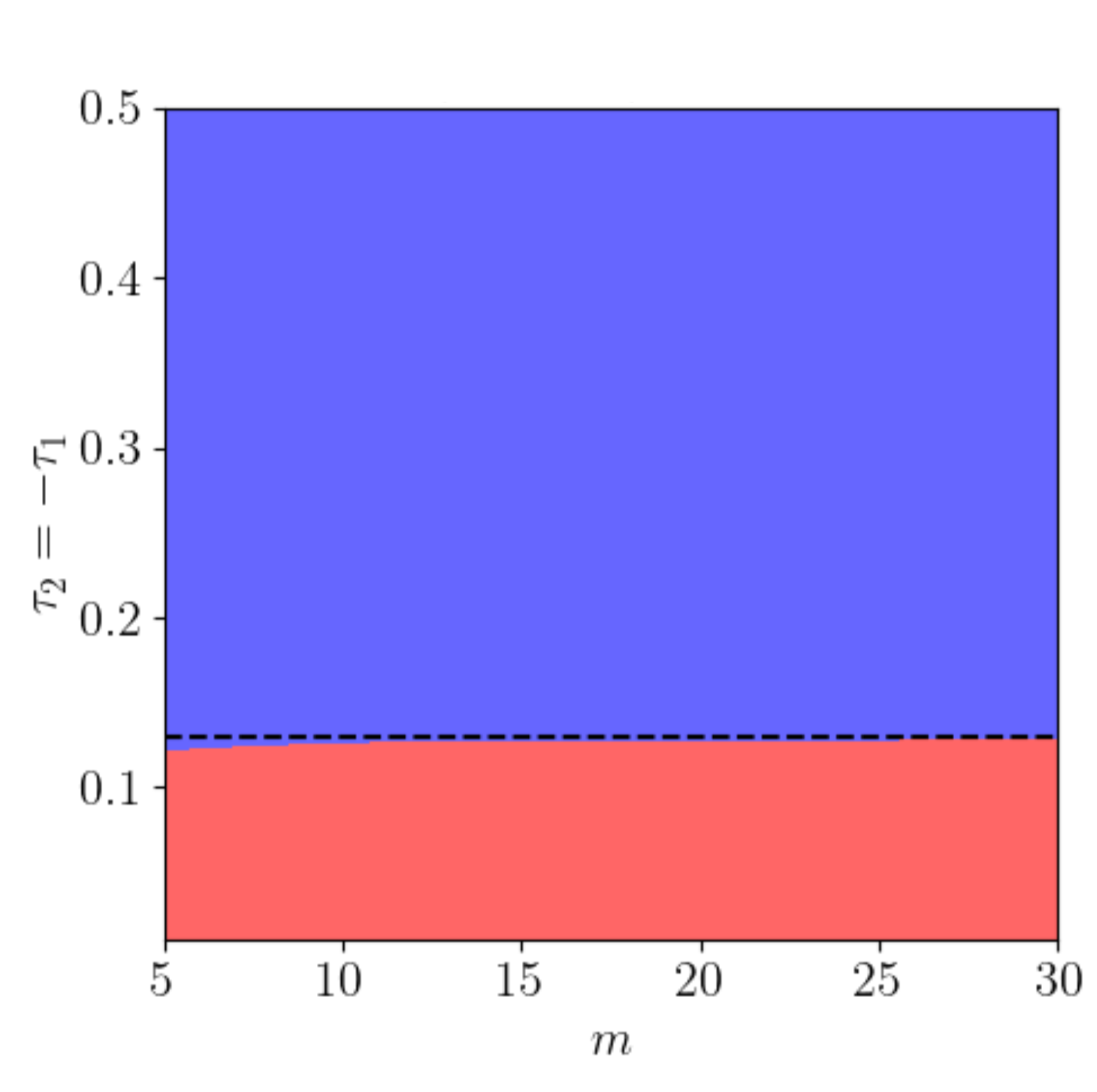}
        \caption{$\btr = 1$, $\btl = 1$}
        \label{fig:temp-ratio-one}
    \end{subfigure}%
    \hfill 
    \begin{subfigure}[b]{.33\textwidth}
        \centering
        \includegraphics[scale=.2]{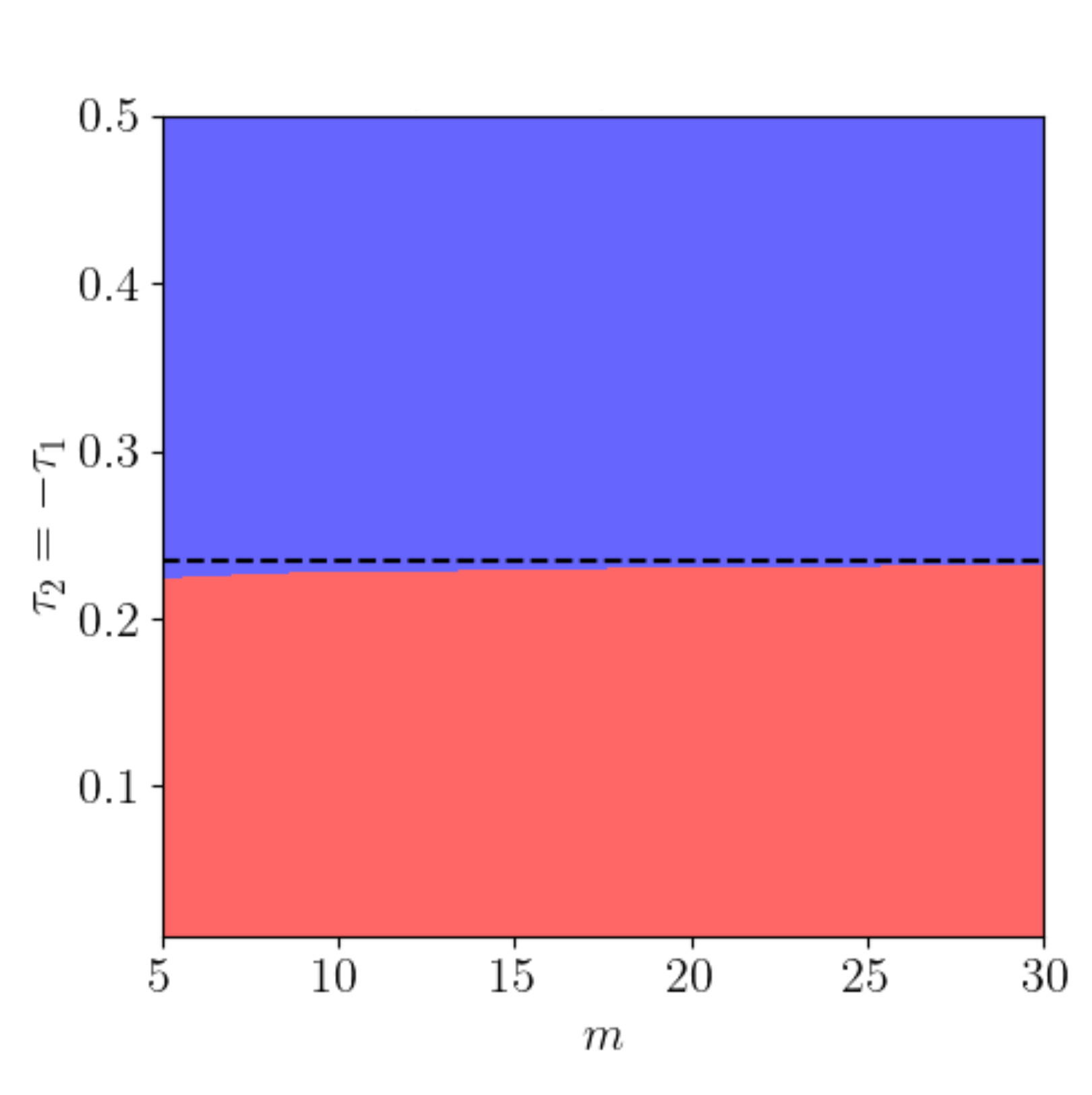}
        \caption{$\btr = 1$, $\btl = 0.2$}
        \label{fig:temp-ratio-five}
    \end{subfigure}%
\caption{The switchover of dominance between the annihilation and propagation saddle is shown as a function of insertion time $\tau_2 = - \tau_1$ and shell mass $m$.   The blue regions indicate the dominance of the annihilation saddle, and the red regions indicate the dominance of the propagation saddle. The black dashed line indicates the switchover value in the large mass limit, which we see the switchover lines here asymptote to. We observe that detection is always possible for sufficiently large $\tau_2$ in the generic mass case.}
    \label{fig:mass-time-plot}
\end{figure}

We would like to understand if our general results in the large mass limit extend to generic interior shell operators  that are sufficiently heavy to backreact on the geometry.
First, we need to solve for the saddle point configuration at arbitrary shell mass. 
Thus, we have to compute the inverse temperatures of the Euclidean disks, i.e. $\hat{\beta}$'s, by solving the saddle point equations in \eqref{eq:grav-propagation-saddle} and \eqref{eq:grav-annihilation-saddle}.
This cannot be done analytically because the shell propagation time $\dtau$  depends implicitly on the inverse temperatures.
Hence, we resort to numerical analysis, which is difficult in arbitrary spacetime dimensions because the implicit $\hat{\beta}$-dependence is more complicated.
Consequently, we work in $(2+1)$-dimensions, where we can write the shell propagation time  \cite{Balasubramanian:2022gmo}
\begin{equation}
    \label{shell-euclidean-time}
    \dtau(\beta_\sR ; \beta_\sL) = \frac{\beta_\sR}{\pi}\arcsin\left(\frac{r_{\sR}}{R_{*}}\right).
\end{equation}
Having determined the inverse temperatures, we can compute the saddle actions and determine saddle dominance depending on the rest mass of the shell. 
Recall from \eqref{eq:two-pt-function} that we are acting with $\psi^\dagger(\tau_2)\psi(\tau_1)$ -- we want $\psi(\tau_1)$ to be inserted close to the shell operator in the Euclidean past, and $\psi^\dagger(\tau_2)$ to be inserted close to the shell operator in the Euclidean future.
Therefore, we choose $\tau_2 = -\tau_1 > 0$ -- the annihilation saddle can be made to dominate for this choice of parameters, as seen in the large mass limit case using \eqref{eq:saddle-diff}.
We determine whether or not we get annihilation saddle dominance for a given rest mass of the interior shell, having taken a sufficiently large value of $\tau_2$.
The exchange of dominance of saddles in this setup is depicted in Fig.~\ref{fig:mass-time-plot}, where we show the switchover times $\tau_2$ for different interior shell masses. 
Here we have fixed the units of mass by setting $G =1$. 
Since we are concerned with masses sufficiently heavy to backreact on the geometry, we will ignore the lower shell mass values. 

We observe that for sufficiently large $\tau_2$ the annihilation saddle dominates.
Additionally, there is an increase in the switchover time $\tau_2$ as we increase the rest mass of the interior shell, which plateaus to the switchover value in the large mass limit.  The physical intuition for the increase in switchover time is that increasing the rest mass of the shell causes it to be located deeper in the interior.
The annihilation channel is unfavorable when this happens, as the interior shell and probe trajectories are increasingly separated. Hence, at larger mass, we need a larger $\tau_2$ value for the annihilation saddle to dominate. Further, similar to the large mass limit case, we observe that fixing $\btl$ while increasing $\btr$ similarly requires us to increase $\tau_2$ for annihilation saddle dominance.
As explained earlier, at smaller $\btl$, the propagation saddle dominates because interior shell operators prefer to connect to each other rather than to the probe operators as needed for the annihilation saddle.
Thus, it gets harder to detect the microstate, and we need a larger $\tau_2$ for the annihilation saddle to dominate.
Evidently, for a generic shell mass in $(2+1)$-dimensions, we can tune our parameters such that the annihilation shell dominates, allowing us to detect the interior shell, or equivalently, the shell microstate.

\subsection{Detecting the microstate from a candidate list}
\label{section:search}

Next, we demonstrate how an observer might detect a particular fixed temperature black hole microstate given a finite list of such microstates where only one matches the target of the search.
First, consider the simplest case in which we are given a list of $N$ such shell states with corresponding operators $\cO_i$ for $i\in\{1,\dots,N\}$ -- we are told that exactly one of them corresponds to the black hole microstate and that there are no overlaps between any of these $N$ states. 
The other states are not required to be shell states in this first instance.
In this scenario, one can perform a simple binary search to detect the microstate.

We choose a probe operator given by a uniform linear combination of the first $\lfloor N/2 \rfloor$ states
\begin{equation}
    \label{eq:easy-binary-search-op}
    \psi_1 = \frac{1}{\sqrt{\lfloor N/2 \rfloor}} \sum_{i=1}^{\lfloor N/2 \rfloor}\cO_i.
\end{equation}
We input this probe operator into our Euclidean two-point function \eqref{eq:two-pt-function} with appropriately chosen Euclidean preparation times.
If we detect a signal from the annihilation saddle, indicating that one of the candidates matches the interior shell operator, then we repeat this search protocol on these first $\lfloor \frac{N}{2} \rfloor$ operators. 
Otherwise, we take the second set of operators and do the same.
Thus, we will find the operator that matches the interior shell operator in $O(\log_2 N)$ steps.
 Note that at each step of this protocol, we need to compute the Euclidean two-point function in the shell microstate being detected.

Now, let us relax the constraint that other operators have no overlap. 
It was shown in \cite{Balasubramanian:2022gmo} that two arbitrary shell states have an exponentially small overlap.
These overlaps will mean that we also get an exponentially small annihilation saddle contribution from each of these operators.
Since these overlaps are exponentially suppressed, the binary search protocol in $O(\log_2 N)$ steps will still be successful as long as $N$ is parametrically smaller than $e^{S_{\BH}}$.

\section{Boundary detection via eigenstate thermalization hypothesis} 
\label{section:ETH}

In this section, we study black hole microstate detection from the perspective of the dual boundary theory. 
Having found the annihilation and propagation saddle in the bulk AdS theory, a natural next question is if we can find the corresponding saddles in the dual CFT.
The current understanding is that the semiclassical gravitational path integral computes some kind of  coarse-grained average over an underlying ensemble, although precisely what this ensemble consists of is not yet settled
\cite{Cotler:2016fpe,Saad:2018bqo,Saad:2019lba,Penington:2019kki, Marolf:2020xie, Chandra:2022bqq}. In the AdS/CFT correspondence, there is evidence that the coarse-graining in question occurs, at least for relatively simple probe  operators, because  CFT states dual to heavy AdS configurations should satisfy the Eigenstate Thermalization Hypothesis (ETH).  ETH provides a  coarse-grained statistical description that takes a universal form \cite{Sasieta:2022ksu,Balasubramanian:2022gmo, deBoer:2023vsm, Stanford:2020wkf,Altland:2020ccq,Collier:2019weq,Cotler:2021cqa}:  the expectation values of operators in the dual CFT are given by a diagonal matrix comprised of the microcanonical expectation values plus corrections that are exponentially suppressed.
This idea has been studied in great detail in JT gravity \cite{Saad:2019pqd,Jafferis:2022uhu,Jafferis:2022wez}.
The ETH ansatz states that  matrix elements of any operator in the energy basis takes the form\cite{Deutsch:1991msp,Srednicki:1994mfb,Srednicki:1999bhx} \footnote{See also \cite{Pollack:2020gfa} for a another approach.} 
\begin{equation}
\label{eth-ansatz}
    \bra{a} \psi \ket{b} = \overline{\psi} (E_a) \delta_{ab} + e^{-f^{\psi}(E_a, E_b)/2} R^{\psi}_{ab}, 
\end{equation}
where $\overline{\psi} (E_a)$ is the average value of the operator in the microcanonical energy band and $f^{\psi}(E_a, E_b)$ characterizes the variance. 
Both of these functions are expected to be smooth and depend on the details of the operator and the system. 
$R_{ab} \sim \mathcal{N}(0,1)$ is a complex Gaussian random matrix  that gives a statistical interpretation to the corrections.
This ETH ansatz has been extensively used to study the ensemble averages that are computed by the gravitational path integral \cite{Balasubramanian:2022gmo, Balasubramanian:2022lnw, Belin:2020jxr, deBoer:2024mqg, Geng:2025efs, Chandra:2023dgq}.

Recall that the fixed temperature black hole microstate is given by
\begin{equation}
\label{eq:bh-microstate-eth}
   \ket{\Psi} 
    = \ket{\rho_{\btl/2} \cO \rho_{\btr/2}}
    = \frac{1}{\sqrt{Z_1}} \sum_{a,b}e^{-\frac{1}{2}(\btl E_a + \btr E_b)}\cO_{ab}\ket{a,b} \, .
\end{equation}
Using \eqref{eth-ansatz} for the operator $\cO$, this state can be written as
\begin{equation}
    \ket{\Psi} = \frac{1}{\sqrt{Z_1}}\sum_{a,b}e^{-\frac{1}{2}(\btl E_a + \btr E_b - f^{\cO}(E_a, E_b))}R_{ab}\ket{a,b},
\end{equation}
where the normalization is given by
\begin{equation} \label{eq:z1-norm}
    Z_1 = \text{tr} \left( \cOd e^{-\btl H}\cO e^{-\btr H} \right) \approx \sum_{a,b} e^{-\btl E_a - \btr E_b - f^{\cO}(E_a, E_b)}.
\end{equation}
Here we used the average value $\overline{\cO}(E) = 0$ which is required for consistency with the gravitational calculations \cite{Sasieta:2022ksu}.
Having prepared our states in this way, we can construct the density matrix representation for our pure state as
\begin{equation}
    \rho = \ket{\Psi}\bra{\Psi} ,
\end{equation}
which enables us to define the reduced density matrices on the left and right boundary CFTs by tracing out the complementary subsystem as 
\begin{align} 
    \label{right-red-density-matrix}
    \rho_R &= \frac{1}{Z_1}e^{-\frac{\btr}{2}H}\cO e^{-\btl H} \cOd e^{-\frac{\btr}{2}H}.
\end{align}

We are now ready to compute the two-point functions for the boundary CFTs.
We will insert the probe operators on the right, so the relevant two-point function is
\begin{equation}
    \begin{split}
    \langle\psi(\tau_1)\psi^{\dagger}(\tau_2)\rangle_{\rho_R}
    &= \text{tr}(\psi(\tau_1)\psi^{\dagger}(\tau_2)\rho_R) \\
    = \frac{1}{Z_1}\text{tr}&\left( e^{(\tau_1 - \btr/2)H}\psi e^{(\tau_2 - \tau_1)H}\psi^{\dagger} e^{-(\tau_2 + \btr/2)H}\cO e^{-\btl H}\cOd \right) \, ,
    \end{split}
\end{equation} 
Similar to the gravity computation, we define the numerator as 
\begin{equation}
    \cT_1 = \text{tr}\left( e^{(\tau_1 - \btr/2)H}\psi e^{(\tau_2 - \tau_1)H}\psi^{\dagger} e^{-(\tau_2 + \btr/2)H}\cO e^{-\btl H}\cOd \right).
\end{equation}

We can evaluate these terms by taking the continuum limit under which the energy difference between adjacent states is taken to zero, so the sum over energy eigenstates in the trace gets replaced by an integral $ \sum_a \to \int \d E_a e^{S(E_a)}$.
In this limit, the normalization in \eqref{eq:z1-norm} becomes
\begin{equation}
    Z_1 
    = \int \d E_a \d E_b \, e^{S(E_a) + S(E_b)}e^{-(\btr E_a + \btl E_b) - f^{\cO}(E_a, E_b)} \, .
\end{equation}
To evaluate these integrals, we use saddle point analysis.
The saddle point equations are 
\begin{equation}
\label{eq:saddle-point-z1}
    S'(E_a) = \btr + \partial_{E_a}f^{\cO}(E_a, E_b) \, , 
    \qquad
    S'(E_b) = \btl + \partial_{E_b}f^{\cO}(E_a, E_b) \, .
\end{equation} 
Note that these equations match with the gravitational saddle point equations in \eqref{eq:prep-time} if we identify $f^{\cO}(E_a, E_b)$ appropriately.
Explicitly, we have
\begin{equation}
\label{eq:envelope-function}
   f^{\cO}(E_a,E_b) = \frac{\dtau(\beta(E_a);\beta(E_b))}{\beta(E_a)}S(E_a) + \frac{\dtau(\beta(E_b);\beta(E_a))}{\beta(E_b)}S(E_b) + \Is(\beta(E_a),\beta(E_b)).
\end{equation}
where the functional forms for $\dtau$ and $\Is$ are given in \eqref{eq:preparation-time} and \eqref{eq:shell-action} respectively.
The conformal dimension of a $\text{CFT}_d$ boundary operator is related to the shell mass in the bulk geometry via
\begin{equation}
    \label{eq:conformal-dim}
    \Delta = \frac{d}{2} + \sqrt{m^2 + \frac{d^2}{4}}.
\end{equation}
The large mass limit on the gravity side corresponds to the limit of large conformal dimension, and $\Delta \approx m$ in this limit. The gravitational results can be recovered from these expressions by choosing $f^{\mathcal O}$ to match the corresponding quantities computed on the bulk side. In the limit of large conformal dimension, this expression simplifies to 
\begin{equation}
\label{eq:envelope-func-large-mass}
    f^{\cO}(E_a, E_b) \approx \Is (m) = 2 \Delta_{\cO} \log R_* \, ,
\end{equation}
because the shell propagation times vanish, and the shell action becomes independent of the inverse temperatures as in \eqref{eq:shell-action-large-mass}.
In the remainder of this section, we will be agnostic to the particular functional form of the $f^{\cO}$-function.

After choosing the appropriate $f^{\cO}$-function, the saddle point equations in \eqref{eq:saddle-point-z1} have the simultaneous solutions $E_a = M_R$ and $E_b = M_L$, 
corresponding to the masses of the left and right black holes, respectively. Also, we can identify the inverse temperatures of these black holes as
\begin{equation}
    \beta_{\sL} = S'(M_{\sL}) \, , 
    \quad\quad
    \beta_{\sR} = S'(M_{\sR}) \, . 
\end{equation}
This concludes the analysis for the normalization $Z_1$.

\subsection{Light Operators}

We first consider light probe operators.
Recall that from the bulk perspective, these light probes do not backreact on the geometry, so the gravitational saddle is unchanged.  Thus, these probes should not allow us to detect the microstate.  In this subsection, we confirm this by performing an ETH calculation in the dual CFT.

Employing the ETH ansatz in \eqref{eth-ansatz}, 
\begin{equation}
    \label{eq:eth-light}
    \begin{aligned}
        \cO_{ab} & = \bra{a}\cO\ket{b} \approx e^{-f^{\cO}(E_a, E_b)/2}R^{\cO}_{ab} \, , \\
        \psi_{ab} & = \bra{a}\psi\ket{b} \approx \overline{\psi}(E_a)\delta_{ab} + e^{-f^{\psi}(E_a, E_b)/2}R^{\psi}_{ab} \, ,
    \end{aligned}
\end{equation}
where for the operator corresponding to the black hole shell microstate $\cO$, we assume $\overline{\cO}(E_a)=0$.
This can be achieved by redefining the operator with the average value subtracted out. 
Note that this  $\overline{\cO}(E_a) = 0$ condition is imperative if we want the ETH results to match the gravity results \cite{Sasieta:2022ksu, Balasubramanian:2022gmo}.
It is also possible to choose $\overline{\psi}(E_a) = 0$, but we will keep it arbitrary.

Since the operators $\psi$ and $\cO$ are uncorrelated, we can assume that the random matrices $R^{\psi}$ and $R^{\cO}$ are  independent.
Under this assumption,  the numerator term becomes 
\begin{multline} 
\label{eq:trace-probe-eth}
    \cT^{\psi}_1 
    = \int \d E_a \d E_d e^{S(E_a) + S(E_d)} e^{- (\btr E_a + \btl E_d) - f^{\cO} (E_a, E_d)} \left| \overline{\psi} (E_a) \right|^2 \\
    + \int \d E_a \d E_b \d E_d e^{S(E_a) + S(E_b) + S(E_d)} e^{- (\btr + \tau_1 - \tau_2)E_a - (\tau_2 - \tau_1)E_b - \btl E_d - f^{\psi} (E_a, E_b) - f^{\cO} (E_a, E_d) }.
\end{multline}
We now take the limit in which the probe operator is light, which means that its conformal dimension is $\Delta_{\psi} = d + \epsilon$ with $\epsilon$ small.
In this limit, we can use saddle-point analysis to obtain
\begin{equation} 
    \cT^{\psi}_1 
    \approx Z_1 \times \left( \left| \overline{\psi} (M_R) \right|^2 + \int \d E_b \, e^{S(E_b) - (\tau_2 - \tau_1)(M_R - E_b) - f^{\psi}(M_R,E_b)} \right)  \, .
\end{equation}
This is because the light operator must not affect the saddle-point configuration.
Thus, the Euclidean correlator evaluates to
\begin{equation}
\label{eq:corr-light-probe}
    \langle \psi(\tau_1)\psi^{\dagger}(\tau_2) \rangle_{\rho_R} = \left| \overline{\psi} (M_R) \right|^2 + \int \d E_b \, e^{S(E_b) - (\tau_2 - \tau_1)(M_R - E_b) - f^{\psi}(M_R,E_b)} \, . 
\end{equation}

We can compare this result to the thermal two point function
\begin{equation}
    \langle \psi(\tau_1) \psi^{\dagger}(\tau_2) \rangle_{\text{th}}
    = \frac{\cT_0}{Z_0} 
    = \frac{1}{Z_0} \text{tr} \left( e^{(\tau_1 - \tau_2 - \btr) H}\psi e^{(\tau_2 - \tau_1) H} \psi^{\dagger} \right),
\end{equation}
where the normalization is
\begin{equation}
    Z_0 = \text{tr} \left( e^{-\btr H} \right) \, .
\end{equation}
Using a similar analysis as above, we obtain that for a light probe operator $\psi$
\begin{equation} \label{eq:trace-thermal-eth}
    \cT_0 = \int \d E_a e^{S(E_a)} e^{-\br E_a} \left| \overline{\psi} (E_a) \right|^2 + \int \d E_a \d E_b e^{S(E_a) + S(E_b)} e^{-(\btr + \tau_1 - \tau_2)E_a - (\tau_2 - \tau_1)E_b - f^{\psi}(E_a,E_b)} \, ,
\end{equation}
and
\begin{equation}
\label{eq:z0}
    Z_0 = \int \d E_a \,  e^{S(E_a) - \br E_a} \, .
\end{equation}
Taking the probe operator to be light, it follows that
\begin{equation}
    \cT_0 \approx Z_0 \times \left( \left| \overline{\psi} (M_R) \right|^2 + \int \d E_b \, e^{S(E_b) - (\tau_2 - \tau_1)(M_R - E_b) - f^{\psi}(M_R,E_b)} \right) \, ,
\end{equation}
so the thermal two-point function is
\begin{equation}
    \langle \psi(\tau_1) \psi^{\dagger}(\tau_2) \rangle_{\text{th}} = \left| \overline{\psi} (M_R) \right|^2 + \int \d E_b \, e^{S(E_b) - (\tau_2 - \tau_1)(M_R - E_b) - f^{\psi}(M_R,E_b)} \, . 
\end{equation}
This matches the result in \eqref{eq:corr-light-probe}.
Evidently, it is impossible to distinguish between the black hole microstate and the thermal state by using the Euclidean two-point function for a light probe.

\subsection{Heavy Operators}

In this subsection, we consider the case of probe operators with large conformal dimension corresponding to massive shells that significantly backreact on the dual bulk geometry. 
Thus, they should have a non-trivial impact on the saddle point, unlike the previous case of light probe operators. 

First, we consider a heavy probe operator $\psi = \phi$ that is orthogonal to $\cO$, the interior shell operator that prepares the black hole microstate \eqref{eq:bh-microstate-eth}.
In this case, the  ETH ansatz takes the following form 
\begin{equation}
\label{eq:eth-heavy}    
\begin{aligned}
    \cO_{ab} & = \bra{a}\cO\ket{b} \approx e^{-f^{\cO}(E_a, E_b)/2}R^{\cO}_{ab} \, , \\
    \phi_{ab} & = \bra{a}\phi\ket{b} \approx e^{-f^{\phi}(E_a, E_b)/2}R^{\phi}_{ab} \, ,
\end{aligned}
\end{equation}
 where we assume $\overline{\cO}(E_a) = \overline{\phi_i}(E_a) = 0$ as explained earlier.
Note that the orthogonality condition above concretely means that the random matrices $R^{\phi}$ and $R^{\cO}$ are completely independent.

The numerator of the two-point function for this heavy orthogonal probe is
\begin{equation} \label{eq:orthogonal-numerator}
    \cT^{\phi}_1
    = \int \d E_a \d E_b \d E_d \, e^{S(E_a) + S(E_b) + S(E_d)} e^{-(\btr E_a + \btl E_d) + (\tau_2 - \tau_1)(E_a - E_b) - f^{\cO}(E_a, E_d)}  e^{-f^{\phi}(E_a, E_b)} \, .
\end{equation}
As earlier, we can evaluate this using a saddle point analysis.
The saddle point equations are
\begin{equation}
\label{eq:propagation-saddle-eth}
\begin{split}
    S' (E_a)
    &= \btr + \tau_1 - \tau_2 + \partial_{E_a}f^{\cO}(E_a, E_d) + \partial_{E_a}f^{\phi}(E_a, E_b) \, ,  \\
    S' (E_b) 
    &= \tau_2 - \tau_1 + \partial_{E_b}f^{\phi}(E_a, E_b) \, , \\ 
    S' (E_d) 
    &= \btl + \partial_{E_d}f^{\cO}(E_a, E_d) \, .
\end{split}
\end{equation}
These equations correspond to the propagation saddle in the gravitational analysis.
Indeed, we exactly recover the gravitational saddle point equations in \eqref{eq:grav-propagation-saddle} after choosing $f^{\phi}$ and $f^{\cO}$  in \eqref{eq:envelope-function}.

Next, we take the probe operator to be the interior shell operator $\psi = \cO$.
Again, we compute the numerator of the two-point function, which now has two copies each of $R^{\cO}$ and $\left(R^{\cO}\right)^\dagger$. 
Since we can contract these in two different ways, the numerator is given by a sum of two terms 
\begin{equation} \label{eq:probe-matched-eth}
\begin{split}
    \cT_1^{\cO}
    =& \,\int \d E_a \d E_b \d E_d \, e^{S(E_a) + S(E_b) + S(E_d)} e^{-(\btr E_a + \btl E_d) + (\tau_2 - \tau_1)(E_a - E_b) - f^{\cO}(E_a, E_d)} e^{-f^{\cO}(E_a, E_b)} \\
    & + \int \d E_a \d E_b \d E_c \, e^{S(E_a) + S(E_b) + S(E_c)} e^{- \btr E_a/2 + \btl E_b + \btr E_c/2 + \tau_1 (E_b - E_a) + \tau_2 (E_c - E_b) } \\
    & \hspace{4in} \times e^{- f^{\cO}(E_b, E_c) - f^{\cO}(E_a, E_b)} \, .
\end{split}
\end{equation}
The first term corresponds to the propagation saddle because the saddle point equations are identical to \eqref{eq:propagation-saddle-eth}.
The saddle point equations for the second term are
\begin{equation}
\label{eq:annihilation-saddle-eth}
\begin{split}
    S' (E_a) 
    &= \frac{1}{2} \btr + \tau_1 + \partial_{E_a} f^{\cO}(E_a, E_b) \, , \\
    S' (E_b)
    &= \btl + \tau_2 - \tau_1 + \partial_{E_b}f^{\cO}(E_b, E_c) + \partial_{E_b}f^{\cO}(E_a, E_b) \, , \\    
    S' (E_c) 
    & = \frac{1}{2} \btr - \tau_2 + \partial_{E_c}f^{\cO}(E_b, E_c) \, .
\end{split}
\end{equation}
These equations correspond to the propagation saddle in the gravitational analysis.
As earlier, we recover the gravitational saddle point equations in \eqref{eq:grav-annihilation-saddle} after choosing $f^{\cO}$ appropriately.

Lastly, we can decompose an arbitrary probe as a linear combination of orthonormal operators $\{ \cO, \phi_1, \phi_2, \dots \}$ as in \eqref{eq:op-complex}
\begin{equation}
\label{eq:op-complex-eth}
    \psi = w_{\cO} \cO + \sum_{i} w_i \phi_i \, .
\end{equation}
Substituting the ETH ansatz in \eqref{eq:eth-heavy} for each operator on the right side, we have 
\begin{equation}
    \psi_{ab} = \bra{a}\psi\ket{b} \approx w_{\cO} e^{-f^{\cO}(E_a, E_b)/2}R^{\cO}_{ab} + \sum_{i}  w_i e^{-f^{\phi_i}(E_a, E_b)/2}R^{\phi_i}_{ab} \, ,
\end{equation}
so the numerator term evaluates to 
\begin{equation}
\label{eq:blackhole-trace-num}
\begin{split}
    \cT_1
    = \, & w_{\cO}^2 \int \d E_a \d E_b \d E_d \, e^{S(E_a) + S(E_b) + S(E_d)} e^{-(\btr E_a + \btl E_d) + (\tau_2 - \tau_1)(E_a - E_b) - f^{\cO}(E_a, E_d)} e^{-f^{\cO}(E_a, E_b)} \\
    &+ w_{\cO}^2 \int \d E_a \d E_b \d E_c \, e^{S(E_a) + S(E_b) + S(E_c)} e^{- \btr E_a/2 + \btl E_b + \btr E_c/2 + \tau_1 (E_b - E_a) + \tau_2 (E_c - E_b)} \\
    & \hspace{4in} \times e^{- f^{\cO}(E_b, E_c) - f^{\cO}(E_a, E_b)} \\
    &+ \sum_{i} w_i^2 \int \d E_a \d E_b \d E_d \, e^{S(E_a) + S(E_b) + S(E_d)} e^{-(\btr E_a + \btl E_d) + (\tau_2 - \tau_1)(E_a - E_b) - f^{\cO}(E_a, E_d)}  e^{-f^{\phi_i}(E_a, E_b)} \, .
    \end{split}
\end{equation}
Indeed, the first and third terms correspond to the propagation saddle, which exists for all operators in the decomposition in \eqref{eq:op-complex-eth}.
The second term corresponds to the annihilation saddle and only exists for the operator $\cO$.
This is the term that leads to detectability.
From these results, we can conclude that the ETH answers perfectly match the gravitational path integral calculations.

\section{Variance in the Euclidean correlator from ensemble averaging}
\label{section:noise}

This gravitational path integral is believed to compute an average over an underlying ensemble, which we have modeled in the dual CFT in terms of  ETH. 
One natural concern is that the results for a particular member of the ensemble may be significantly different from the average value, and the dominance of the annihilation saddle could be overpowered by  statistical fluctuations.
We can characterize these statistical effects in terms of the variance of the numerator term $\sigma_{\cT_1}^2$.
In this section, we show that the annihilation saddle can be made to dominate over these statistical effects by tuning the parameters $\tau_1$, $\tau_2$. 

First, we compute the variance from the boundary perspective by using ETH.
The squared modulus of the numerator term for an arbitrary probe operator, as in \eqref{eq:op-complex-eth}, is given by
\begin{multline}
\label{eq:subleading}
    |\cT_1|^2 
    = \sum_{a,b,c,d} \sum_{a',b',c',d'} 
    e^{-(\btr/2 + \tau_1) (E_a + E_{a'}) - (\tau_2 - \tau_1) (E_b + E_{b'}) - (\btr/2 - \tau_2) (E_c + E_{c'}) - \btl (E_d + E_{d'})} \\
    \times \bigg( w_{\cO} \cO_{ab} + \sum_{i} w_{i} \big( \phi_{i} \big)_{ab} \bigg) \Big( w_{\cO} \cO^{*}_{cb} + \sum_{j} w_{j} \big( \phi_{j} \big)^{*}_{cb} \bigg) \cO_{cd} \cO^{*}_{ad} \\
    \times \bigg( w_{\cO} \cO^{*}_{a'b'} + \sum_{i'}  w_{i'} \big( \phi_{i'} \big)^{*}_{a'b'} \bigg) \bigg( w_{\cO} \cO_{c'b'} + \sum_{j'} w_{j'} \big( \phi_{j'} \big)_{c'b'} \bigg) \cO^{*}_{c'd'} \cO_{a'd'} \, ,
\end{multline} 
We can evaluate this expression in terms of the ETH ansatz in \eqref{eq:eth-heavy} by performing the contractions between the random matrices $R$ and $R^\dagger$.
The variance only receives contributions from terms where we contract at least some of unprimed indices with primed indices.  This is because the terms where the unprimed and primed indices are respectively contracted within themselves cancel when we subtract out $\cT_1 \times \cT_1^*$.

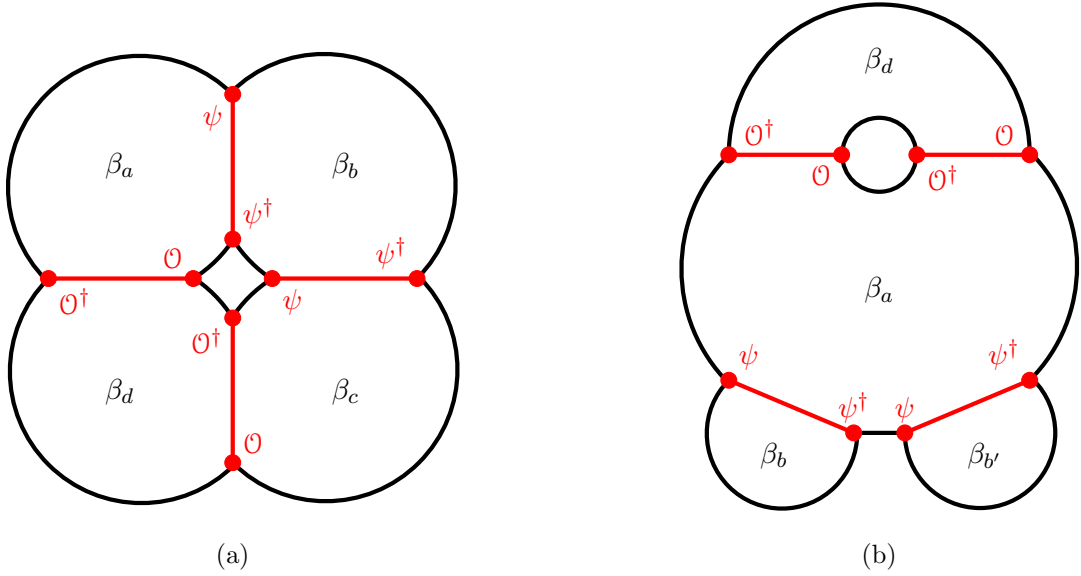
\begin{figure}[t]
    \begin{subfigure}{.5\textwidth}
        \centering
        \begin{tikzpicture}
            \draw[ultra thick] (0,2.45) arc (45:225:1.75);
            \draw[ultra thick] (-2.45,0) arc (135:315:1.75);
            \draw[ultra thick] (0,-2.45) arc (225:405:1.75);
            \draw[ultra thick] (2.45,0) arc (315:495:1.75);
            
            \draw[ultra thick] (0,0.525) arc (212:238:1.75);
            \draw[ultra thick] (-0.525,0) arc (302:328:1.75);
            \draw[ultra thick] (0,-0.525) arc (32:58:1.75);
            \draw[ultra thick] (0.525,0) arc (122:148:1.75);
    
            \draw[ultra thick,red] (0,0.525) -- (0,2.45);
            \draw[ultra thick,red] (-0.525,0) -- (-2.45,0);
            \draw[ultra thick,red] (0,-0.525) -- (0,-2.45);
            \draw[ultra thick,red] (0.525,0) -- (2.45,0);
    
            \filldraw[red] (0,2.45) circle (3pt) node[anchor=north east]{$\psi$};
            \filldraw[red] (-2.45,0) circle (3pt) node[anchor=north west]{$\cO^{\dagger}$};
            \filldraw[red] (0,-2.45) circle (3pt) node[anchor=south west]{$\cO$};
            \filldraw[red] (2.45,0) circle (3pt) node[anchor=south east]{$\psi^{\dagger}$};
            \filldraw[red] (0,0.525) circle (3pt) node[anchor=south west]{$\psi^{\dagger}$};
            \filldraw[red] (-0.525,0) circle (3pt) node[anchor=south east]{$\cO$};
            \filldraw[red] (0,-0.525) circle (3pt) node[anchor=north east]{$\cO^{\dagger}$};
            \filldraw[red] (0.525,0) circle (3pt) node[anchor=north west]{$\psi$};
    
            \draw (-1.5,1.5) node{$\beta_a$};
            \draw (1.5,1.5) node{$\beta_b$};
            \draw (1.5,-1.5) node{$\beta_c$};
            \draw (-1.5,-1.5) node{$\beta_d$};
        \end{tikzpicture}            
        \caption{}
        \label{fig:sub-dominant-noise}
    \end{subfigure}%
    \hfill
    \begin{subfigure}{.5\textwidth}
        \centering
        \begin{tikzpicture}
            \draw[ultra thick] (0,0) circle (14pt);
            \draw[ultra thick] (2,0) arc (0:180:2);
            \draw[ultra thick] (2,-3) arc (315:405:2.15);
            \draw[ultra thick] (-2,0) arc (135:225:2.15);
            \draw[ultra thick] (-2,-3) arc (135:360:1);
            \draw[ultra thick] (-0.34,-3.7) -- (0.34,-3.7);
            \draw[ultra thick] (0.34,-3.7) arc (180:405:1);
    
            \draw[ultra thick,red] (2,0) -- (0.5,0);
            \draw[ultra thick,red] (-2,0) -- (-0.5,0);
            \draw[ultra thick,red] (2,-3) -- (0.34,-3.7);
            \draw[ultra thick,red] (-2,-3) -- (-0.34,-3.7);

            \filldraw[red] (2,0) circle (3pt) node[anchor= south east]{$\cO\,$};
            \filldraw[red] (-2,0) circle (3pt) node[anchor=south west]{$\,\cO^{\dagger}$};
            \filldraw[red] (0.5,0) circle (3pt) node[anchor=north west]{$\cO^{\dagger}$};
            \filldraw[red] (-0.5,0) circle (3pt) node[anchor=north east]{$\cO$};
            \filldraw[red] (2,-3) circle (3pt) node[anchor=south east]{$\psi^{\dagger}$};
            \filldraw[red] (0.34,-3.7) circle (3pt) node[anchor=south]{$\psi$};
            \filldraw[red] (-0.34,-3.7) circle (3pt) node[anchor=south]{$\psi^{\dagger}$};
            \filldraw[red] (-2,-3) circle (3pt) node[anchor=south west]{$\psi$};

            \draw (0,1.25) node{$\beta_d$};
            \draw (0,-1.8) node{$\beta_a$};
            \draw (-1.4,-4) node{$\beta_b$};
            \draw (1.4,-4) node{$\beta_{b'}$};
        \end{tikzpicture}   
        \caption{}
        \label{fig:dominant-noise}
    \end{subfigure}%
    \caption{Two diagrams that contribute to the variance $\sigma_{\cT_1}^2$.}
    \label{fig:noise-saddles}
\end{figure}

We work in the limit of large conformal dimension to simplify our computations, so we can use \eqref{eq:envelope-func-large-mass}. As in the previous section, we take the continuum limit for the energy, such that the energy difference between adjacent eigenstates is taken to zero, and then we evaluate the integrals using saddle point analysis. After an explicit computation, we obtain the variance
\begin{multline}
\label{eq:variance-2}
    \sigma_{\cT_1}^2 \approx \bigg[ Z \left( \btr + 2 \tau_1 \right) Z \left(\btr - 2 \tau_2 \right) Z\left(2\tau_2 - 2\tau_1 \right) Z \left(2 \btl \right)  \\
    + Z \left(2\btr - 2\tau_2 + 2\tau_1 \right) Z \left(\tau_2 - \tau_1 \right)^2 Z \left(2 \btl \right) + Z \left(2\btr - 2\tau_2 + 2\tau_1 \right) Z \left(2\tau_2 - 2\tau_1 \right) Z \left( \btl \right)^2  \\
    + 2 w_{\cO}^2 \, Z \left( 3\btr/2 - 2\tau_2 + \tau_1 \right)  Z \left( \btr/2 + \tau_1 \right) Z \left( 2\btl + \tau_2 - \tau_1 \right) Z \left( \tau_2 - \tau_1 \right) \\
    + 2 w_{\cO}^2 \, Z \left( 3\btr/2 - \tau_2 + 2 \tau_1 \right) Z \left(\btr/2 - \tau_2 \right) Z \left(2\btl + \tau_2 - \tau_1 \right) Z \left( \tau_2 - \tau_1 \right) \\
    + 2 w_{\cO}^2 \, Z \left( 3\btr/2 - 2\tau_2 + \tau_1 \right)  Z \left( \btr/2 + \tau_1 \right) Z \left( \btl + 2\tau_2 - 2\tau_1 \right) Z \left( \btl \right) \\
    + 2 w_{\cO}^2 \, Z \left( 3\btr/2 - \tau_2 + 2 \tau_1 \right) Z \left(\btr/2 - \tau_2 \right) Z \left( \btl + 2\tau_2 - 2\tau_1 \right) Z \left( \btl \right) \\
    + 2 w_{\cO}^2 \, Z \left( 2\btr - 2\tau_2 + 2\tau_1 \right) Z \left(\btl + \tau_2 - \tau_1 \right) Z \left(\tau_2 - \tau_1 \right) Z \left( \btl \right) \\
    + w_{\cO}^2 \, Z \left(\btr + 2 \tau_1 \right) Z \left( \btr/2 - \tau_2 \right)^2 Z \left(2\btl + 2\tau_2 - 2\tau_1 \right) \\
    + w_{\cO}^2 \, Z \left(\btr/2 + \tau_1 \right)^2 Z \left( \btr - 2\tau_2 \right) Z \left(2\btl + 2\tau_2 - 2\tau_1 \right) \\
    + w_{\cO}^4 \, Z \left( \btr - \tau_2 + \tau_1 \right)^2 Z \left(\btl + \tau_2 - \tau_1 \right)^2 \\
    + 2 w_{\cO}^4 \, Z \left( \btr - \tau_2 + \tau_1 \right) Z \left(\btr/2 + \tau_1 \right) Z \left(\btr/2 - \tau_2 \right) Z \left(2\btl + 2\tau_2 - 2\tau_1 \right) \\
    + 2 w_{\cO}^2 \, Z \left( 2\btr - 2\tau_2 + 2\tau_1 \right) Z \left(2\btl + 2\tau_2 - 2\tau_1 \right) \bigg] \times e^{- 4 \Is(m)} \, .
\end{multline}
Note that each of these terms is in a one-to-one correspondence with a diagram contributing to the gravitational path integral.
For instance, the diagrams corresponding to the first two terms are shown in Fig.~\ref{fig:noise-saddles}.
Although we have used the ETH analysis here, we could also have computed the variance directly using the gravitational path integral and drawing all relevant diagrams.

We can now compare the variance $\sigma_{\cT_1}^2$ to the annihilation saddle contribution in \eqref{eq:saddle-disk-annihilation}\footnote{We have included the factor of $w_{\cO}^2$ in this expression.}
\begin{equation}
    \cT^{A}_1 = w_{\cO}^2 Z \left( \btr/2 - \tau_2 \right) Z \left( \btl + \tau_2 - \tau_1 \right)  Z \left( \btr/2 + \tau_1 \right) e^{- 2 \Is(m)} \, .
\end{equation}
Recall that the annihilation contribution becomes large when the probe operators are inserted close to the interior shell operators, i.e., $\tau_1 \to - \btr/2$ and $\tau_2 \to \btr/2$.
This tuning was crucial to get the annihilation saddle to dominate over the propagation saddle and achieve detectability. 
One can check that in this limit, none of the terms in \eqref{eq:variance-2} grows as rapidly as $(\cT^{A}_1)^2$.
The intuition behind this result is that the variance only receives contributions from connected diagrams.
These diagrams have the topology of either an annulus (Euler characteristic $\chi_E = 0$) or an annulus with a handle ($\chi_E = -2$).
However, $(\cT^{A}_1)^2$ is given by two disconnected disks ($\chi_E = 2$), so it can be made to dominate over the variance.

We can see this explicitly in the particular case of $\tau_1 = - \tau_2$.
In the limit $\tau_2 \to \btr/2$, the dominant contribution to the variance is
\begin{equation}
\label{eq:variance-dominant}
    \sigma_{\cT_1}^2 \supset  
    2 (w_{\cO}^2 + w_{\cO}^4) Z \left(\btr - 2 \tau_2 \right) Z \left( \btr/2 - \tau_2 \right)^2 Z \left( 2\btl + 4\tau_2 \right) e^{- 4 \Is(m)} \, .
\end{equation}
We can divide this by $(\cT^{A}_1)^2$ to obtain
\begin{equation}
\label{eq:variance-dominant-ratio}
    \frac{\sigma_{\cT_1}^2}{(\cT^{A}_1)^2} \supset  
    \frac{2 + 2 w_{\cO}^2}{w_{\cO}^2} \frac{Z \left(\btr - 2 \tau_2 \right) Z \left( 2\btl + 4\tau_2 \right)}{Z \left( \btr/2 - \tau_2 \right)^2 Z \left( \btl + 2\tau_2 \right)^2} \, .
\end{equation}
Using \eqref{eq:free-energy}, we can see that this term be made small if we choose $\tau_2$ appropriately close to $\btr/2$.
Moreover, this is precisely the tuning needed for the annihilation saddle contribution to dominate over the propagation saddle contribution. 
In conclusion, the statistical variance does not affect black hole microstate detectability if the Euclidean preparation times are chosen appropriately.

\iffalse

\begin{figure}[t]
    \begin{subfigure}{.5\textwidth}
        \centering
        \includegraphics[scale=1]{noise-small-n.png}
        \caption{$n=2$}
        \label{fig:noise-plot-n=2}
    \end{subfigure}%
    \hfill
    \begin{subfigure}{.5\textwidth}
        \centering
        \includegraphics[scale=1]{noise-large-n.png}
        \caption{$n=10^{11}$}
        \label{fig:noise-plot-n-large}
    \end{subfigure}%
    \caption{The switchover of dominance between the annihilation saddle and noise contributions is shown for $\btl=\btr=1$. The blue region displays annihilation saddle dominance, and the red region displays noise dominance. The gray region shows the forbidden region, as before. What is ``n'' here?}
    \label{fig:noise-domination}
\end{figure}

\fi

\section{Discussion}
\label{section:diss}

The results presented in this paper demonstrate that  Euclidean  correlation functions computed at asymptotic infinity can detect the microstate of a black hole in any dimension through nonperturbative gravitational effects. 
Our analysis was restricted to Euclidean quantities that are not  directly detected by  asymptotic Lorentzian observers.
Indeed, the single-sided Lorentzian two-point function does not have an exponentially large annihilation saddle contribution because we can no longer tune the insertion times to be close to a divergence.  That said, previous work has argued that complex states of gravity, including black hole microstates, could be identified in Lorentzian signature by sufficiently precise asymptotic measurements of quantities like the mass \cite{Balasubramanian:2006iw}
or multipole moments \cite{Balasubramanian:2006jt} of the spacetime.  Likewise see \cite{Balasubramanian:2025akx} for a recent analysis of  Lorentzian quantities that can identify the interior state of a black hole.  

In this paper  we only dealt with the two-sided eternal black hole in the canonical ensemble for  AdS spacetimes.  It would be interesting to extend our analysis to asymptoticaly flat space, to single-sided black holes, and to fixed energy states rather than the fixed temperature ones that we considered here.    We have also not addressed what happens if the spacetime is in a superposition of states, one of which we seek to detect.  In fact, because any sufficiently large set of shell states provides a basis \cite{Balasubramanian:2022gmo,Balasubramanian:2025jeu}, we could regard any given microstate as a superposition of a great many others. In that case, the amplitude for any superposition component will be exponentially small in the entropy.  However, we could take a black hole to be in, for example, an equal superposition of two shell states and ask how to interpret a measurement of the state from the point of view of a bulk observer.  Operationally this would presumably require a description in terms of a collapse of the wavefunction behind the horizon \cite{Balasubramanian:2025qoz}.

\paragraph{Acknowledgments.}
We thank Albion Lawrence, Javier Magan, Martin Sasieta, and Tom Yildirim for helpful discussions.  VB, CM, and WC were supported in part by the DOE through DESC0013528 and the QuantISED grant DE-SC0020360. WC was supported in part by the Thouron Award. VB was supported in part by the Eastman Professorship at Balliol College, University of Oxford.

\appendix

\section*{}
\bibliographystyle{apsrev4-1long}
\bibliography{main}

%Merlin.mbs v4.21 2009-07-09.
\begin{thebibliography}{10}%
\makeatletter
\providecommand \@ifxundefined [1]{%
 \ifx #1\undefined \expandafter \@firstoftwo
 \else \expandafter \@secondoftwo
\fi
}%
\providecommand \@ifnum [1]{%
 \ifnum #1\expandafter \@firstoftwo
 \else \expandafter \@secondoftwo
\fi
}%
\providecommand \enquote [1]{``#1''}%
\providecommand \bibnamefont  [1]{#1}%
\providecommand \bibfnamefont [1]{#1}%
\providecommand \citenamefont [1]{#1}%
\providecommand\href[0]{\@sanitize\@href}%
\providecommand\@href[1]{\endgroup\@@startlink{#1}\endgroup\@@href}%
\providecommand\@@href[1]{#1\@@endlink}%
\providecommand \@sanitize [0]{\begingroup\catcode`\&12\catcode`\#12\relax}%
\@ifxundefined \pdfoutput {\@firstoftwo}{%
 \@ifnum{\z@=\pdfoutput}{\@firstoftwo}{\@secondoftwo}%
}{%
 \providecommand\@@startlink[1]{\leavevmode\special{html:<a href="#1">}}%
 \providecommand\@@endlink[0]{\special{html:</a>}}%
}{%
 \providecommand\@@startlink[1]{%
  \leavevmode
  \pdfstartlink
   attr{/Border[0 0 1 ]/H/I/C[0 1 1]}%
   user{/Subtype/Link/A<</Type/Action/S/URI/URI(#1)>>}%
  \relax
 }%
 \providecommand\@@endlink[0]{\pdfendlink}%
}%
\providecommand \url  [0]{\begingroup\@sanitize \@url }%
\providecommand \@url [1]{\endgroup\@href {#1}{\urlprefix}}%
\providecommand \urlprefix [0]{URL }%
\providecommand \Eprint[0]{\href }%
\@ifxundefined \urlstyle {%
  \providecommand \doi [1]{doi:\discretionary{}{}{}#1}%
}{%
  \providecommand \doi [0]{doi:\discretionary{}{}{}\begingroup \urlstyle{rm}\Url }%
}%
\providecommand \doibase [0]{http://dx.doi.org/}%
\providecommand \Doi[1]{\href{\doibase#1}}%
\providecommand \bibAnnote [3]{%
  \BibitemShut{#1}%
  \begin{quotation}\noindent
    \textsc{Key:}\ #2\\\textsc{Annotation:}\ #3%
  \end{quotation}%
}%
\providecommand \bibAnnoteFile [2]{%
  \IfFileExists{#2}{\bibAnnote {#1} {#2} {\input{#2}}}{}%
}%
\providecommand \typeout [0]{\immediate \write \m@ne }%
\providecommand \selectlanguage [0]{\@gobble}%
\providecommand \bibinfo [0]{\@secondoftwo}%
\providecommand \bibfield [0]{\@secondoftwo}%
\providecommand \translation [1]{[#1]}%
\providecommand \BibitemOpen[0]{}%
\providecommand \bibitemStop [0]{}%
\providecommand \bibitemNoStop [0]{.\EOS\space}%
\providecommand \EOS [0]{\spacefactor3000\relax}%
\providecommand \BibitemShut [1]{\csname bibitem#1\endcsname}%
%</preamble>
\bibitem{Balasubramanian:2006iw}%
  \BibitemOpen
  \bibfield{author}{%
  \bibinfo {author} {\bibfnamefont{Vijay}\ \bibnamefont{Balasubramanian}}, \bibinfo {author} {\bibfnamefont{Donald}\ \bibnamefont{Marolf}},\ and\ \bibinfo {author} {\bibfnamefont{Moshe}\ \bibnamefont{Rozali}},\ }%
  \bibfield{title}{%
  \enquote{\bibinfo {title} {{Information Recovery From Black Holes}},}\ }%
  \bibfield{journal}{%
  \Doi{10.1007/s10714-006-0344-8}{\bibinfo {journal} {Gen. Rel. Grav.}}\ }%
  \textbf{\bibinfo {volume} {38}},\ \bibinfo {pages} {1529--1536} (\bibinfo {year} {2006}),\ \Eprint{http://arxiv.org/abs/hep-th/0604045}{arXiv:hep-th/0604045}%
  \bibAnnoteFile{NoStop}{Balasubramanian:2006iw}%
\bibitem{Brown:2019rox}%
  \BibitemOpen
  \bibfield{author}{%
  \bibinfo {author} {\bibfnamefont{Adam~R.}\ \bibnamefont{Brown}}, \bibinfo {author} {\bibfnamefont{Hrant}\ \bibnamefont{Gharibyan}}, \bibinfo {author} {\bibfnamefont{Geoff}\ \bibnamefont{Penington}},\ and\ \bibinfo {author} {\bibfnamefont{Leonard}\ \bibnamefont{Susskind}},\ }%
  \bibfield{title}{%
  \enquote{\bibinfo {title} {{The Python{\textquoteright}s Lunch: geometric obstructions to decoding Hawking radiation}},}\ }%
  \bibfield{journal}{%
  \Doi{10.1007/JHEP08(2020)121}{\bibinfo {journal} {JHEP}}\ }%
  \textbf{\bibinfo {volume} {08}},\ \bibinfo {pages} {121} (\bibinfo {year} {2020}),\ \Eprint{http://arxiv.org/abs/1912.00228}{arXiv:1912.00228 [hep-th]}%
  \bibAnnoteFile{NoStop}{Brown:2019rox}%
\bibitem{Engelhardt:2021qjs}%
  \BibitemOpen
  \bibfield{author}{%
  \bibinfo {author} {\bibfnamefont{Netta}\ \bibnamefont{Engelhardt}}, \bibinfo {author} {\bibfnamefont{Geoff}\ \bibnamefont{Penington}},\ and\ \bibinfo {author} {\bibfnamefont{Arvin}\ \bibnamefont{Shahbazi-Moghaddam}},\ }%
  \bibfield{title}{%
  \enquote{\bibinfo {title} {{Finding pythons in unexpected places}},}\ }%
  \bibfield{journal}{%
  \Doi{10.1088/1361-6382/ac3e75}{\bibinfo {journal} {Class. Quant. Grav.}}\ }%
  \textbf{\bibinfo {volume} {39}},\ \bibinfo {pages} {094002} (\bibinfo {year} {2022}),\ \Eprint{http://arxiv.org/abs/2105.09316}{arXiv:2105.09316 [hep-th]}%
  \bibAnnoteFile{NoStop}{Engelhardt:2021qjs}%
\bibitem{Balasubramanian:2022fiy}%
  \BibitemOpen
  \bibfield{author}{%
  \bibinfo {author} {\bibfnamefont{Vijay}\ \bibnamefont{Balasubramanian}}, \bibinfo {author} {\bibfnamefont{Arjun}\ \bibnamefont{Kar}}, \bibinfo {author} {\bibfnamefont{Cathy}\ \bibnamefont{Li}},\ and\ \bibinfo {author} {\bibfnamefont{Onkar}\ \bibnamefont{Parrikar}},\ }%
  \bibfield{title}{%
  \enquote{\bibinfo {title} {{Quantum error correction in the black hole interior}},}\ }%
  \bibfield{journal}{%
  \Doi{10.1007/JHEP07(2023)189}{\bibinfo {journal} {JHEP}}\ }%
  \textbf{\bibinfo {volume} {07}},\ \bibinfo {pages} {189} (\bibinfo {year} {2023}),\ \Eprint{http://arxiv.org/abs/2203.01961}{arXiv:2203.01961 [hep-th]}%
  \bibAnnoteFile{NoStop}{Balasubramanian:2022fiy}%
\bibitem{Engelhardt:2024hpe}%
  \BibitemOpen
  \bibfield{author}{%
  \bibinfo {author} {\bibfnamefont{Netta}\ \bibnamefont{Engelhardt}}, \bibinfo {author} {\bibfnamefont{{\r{A}}smund}\ \bibnamefont{Folkestad}}, \bibinfo {author} {\bibfnamefont{Adam}\ \bibnamefont{Levine}}, \bibinfo {author} {\bibfnamefont{Evita}\ \bibnamefont{Verheijden}},\ and\ \bibinfo {author} {\bibfnamefont{Lisa}\ \bibnamefont{Yang}},\ }%
  \bibfield{title}{%
  \enquote{\bibinfo {title} {{Cryptographic Censorship}},}\ }%
  \bibfield{journal}{%
  \Doi{10.1007/JHEP01(2025)122}{\bibinfo {journal} {JHEP}}\ }%
  \textbf{\bibinfo {volume} {01}},\ \bibinfo {pages} {122} (\bibinfo {year} {2025}),\ \Eprint{http://arxiv.org/abs/2402.03425}{arXiv:2402.03425 [hep-th]}%
  \bibAnnoteFile{NoStop}{Engelhardt:2024hpe}%
\bibitem{Balasubramanian:2022gmo}%
  \BibitemOpen
  \bibfield{author}{%
  \bibinfo {author} {\bibfnamefont{Vijay}\ \bibnamefont{Balasubramanian}}, \bibinfo {author} {\bibfnamefont{Albion}\ \bibnamefont{Lawrence}}, \bibinfo {author} {\bibfnamefont{Javier~M.}\ \bibnamefont{Magan}},\ and\ \bibinfo {author} {\bibfnamefont{Martin}\ \bibnamefont{Sasieta}},\ }%
  \bibfield{title}{%
  \enquote{\bibinfo {title} {{Microscopic Origin of the Entropy of Black Holes in General Relativity}},}\ }%
  \bibfield{journal}{%
  \Doi{10.1103/PhysRevX.14.011024}{\bibinfo {journal} {Phys. Rev. X}}\ }%
  \textbf{\bibinfo {volume} {14}},\ \bibinfo {pages} {011024} (\bibinfo {year} {2024}),\ \Eprint{http://arxiv.org/abs/2212.02447}{arXiv:2212.02447 [hep-th]}%
  \bibAnnoteFile{NoStop}{Balasubramanian:2022gmo}%
\bibitem{Balasubramanian:2022lnw}%
  \BibitemOpen
  \bibfield{author}{%
  \bibinfo {author} {\bibfnamefont{Vijay}\ \bibnamefont{Balasubramanian}}, \bibinfo {author} {\bibfnamefont{Albion}\ \bibnamefont{Lawrence}}, \bibinfo {author} {\bibfnamefont{Javier~M.}\ \bibnamefont{Magan}},\ and\ \bibinfo {author} {\bibfnamefont{Martin}\ \bibnamefont{Sasieta}},\ }%
  \bibfield{title}{%
  \enquote{\bibinfo {title} {{Microscopic Origin of the Entropy of Astrophysical Black Holes}},}\ }%
  \bibfield{journal}{%
  \Doi{10.1103/PhysRevLett.132.141501}{\bibinfo {journal} {Phys. Rev. Lett.}}\ }%
  \textbf{\bibinfo {volume} {132}},\ \bibinfo {pages} {141501} (\bibinfo {year} {2024}),\ \Eprint{http://arxiv.org/abs/2212.08623}{arXiv:2212.08623 [hep-th]}%
  \bibAnnoteFile{NoStop}{Balasubramanian:2022lnw}%
\bibitem{Climent:2024trz}%
  \BibitemOpen
  \bibfield{author}{%
  \bibinfo {author} {\bibfnamefont{Ana}\ \bibnamefont{Climent}}, \bibinfo {author} {\bibfnamefont{Roberto}\ \bibnamefont{Emparan}}, \bibinfo {author} {\bibfnamefont{Javier~M.}\ \bibnamefont{Magan}}, \bibinfo {author} {\bibfnamefont{Martin}\ \bibnamefont{Sasieta}},\ and\ \bibinfo {author} {\bibfnamefont{Alejandro}\ \bibnamefont{Vilar~L{\'o}pez}},\ }%
  \bibfield{title}{%
  \enquote{\bibinfo {title} {{Universal construction of black hole microstates}},}\ }%
  \bibfield{journal}{%
  \Doi{10.1103/PhysRevD.109.086024}{\bibinfo {journal} {Phys. Rev. D}}\ }%
  \textbf{\bibinfo {volume} {109}},\ \bibinfo {pages} {086024} (\bibinfo {year} {2024}),\ \Eprint{http://arxiv.org/abs/2401.08775}{arXiv:2401.08775 [hep-th]}%
  \bibAnnoteFile{NoStop}{Climent:2024trz}%
\bibitem{Balasubramanian:2025jeu}%
  \BibitemOpen
  \bibfield{author}{%
  \bibinfo {author} {\bibfnamefont{Vijay}\ \bibnamefont{Balasubramanian}}\ and\ \bibinfo {author} {\bibfnamefont{Tom}\ \bibnamefont{Yildirim}},\ }%
  \bibfield{title}{%
  \enquote{\bibinfo {title} {{A Nonperturbative Toolkit for Quantum Gravity}},}\ }%
   (\bibinfo {year} {2025}),\ \Eprint{http://arxiv.org/abs/2504.16986}{arXiv:2504.16986 [hep-th]}%
  \bibAnnoteFile{NoStop}{Balasubramanian:2025jeu}%
\bibitem{Balasubramanian:2025zey}%
  \BibitemOpen
  \bibfield{author}{%
  \bibinfo {author} {\bibfnamefont{Vijay}\ \bibnamefont{Balasubramanian}}\ and\ \bibinfo {author} {\bibfnamefont{Tom}\ \bibnamefont{Yildirim}},\ }%
  \bibfield{title}{%
  \enquote{\bibinfo {title} {{The Nonperturbative Hilbert Space of Quantum Gravity With One Boundary}},}\ }%
   (\bibinfo {year} {2025}),\ \Eprint{http://arxiv.org/abs/2506.04319}{arXiv:2506.04319 [hep-th]}%
  \bibAnnoteFile{NoStop}{Balasubramanian:2025zey}%
\bibitem{He:2025neu}%
  \BibitemOpen
  \bibfield{author}{%
  \bibinfo {author} {\bibfnamefont{Dongming}\ \bibnamefont{He}}, \bibinfo {author} {\bibfnamefont{Juan}\ \bibnamefont{Hernandez}},\ and\ \bibinfo {author} {\bibfnamefont{Maria}\ \bibnamefont{Knysh}},\ }%
  \bibfield{title}{%
  \enquote{\bibinfo {title} {{Quantum corrected black hole microstates and entropy}},}\ }%
   (\bibinfo {month} {9}\ \bibinfo {year} {2025}),\ \Eprint{http://arxiv.org/abs/2510.02997}{arXiv:2510.02997 [hep-th]}%
  \bibAnnoteFile{NoStop}{He:2025neu}%
\bibitem{Balasubramanian:2024rek}%
  \BibitemOpen
  \bibfield{author}{%
  \bibinfo {author} {\bibfnamefont{Vijay}\ \bibnamefont{Balasubramanian}}, \bibinfo {author} {\bibfnamefont{Ben}\ \bibnamefont{Craps}}, \bibinfo {author} {\bibfnamefont{Juan}\ \bibnamefont{Hernandez}}, \bibinfo {author} {\bibfnamefont{Mikhail}\ \bibnamefont{Khramtsov}},\ and\ \bibinfo {author} {\bibfnamefont{Maria}\ \bibnamefont{Knysh}},\ }%
  \bibfield{title}{%
  \enquote{\bibinfo {title} {{Counting microstates of out-of-equilibrium black hole fluctuations}},}\ }%
  \bibfield{journal}{%
  \Doi{10.1007/JHEP06(2025)083}{\bibinfo {journal} {JHEP}}\ }%
  \textbf{\bibinfo {volume} {06}},\ \bibinfo {pages} {083} (\bibinfo {year} {2025}),\ \Eprint{http://arxiv.org/abs/2412.06884}{arXiv:2412.06884 [hep-th]}%
  \bibAnnoteFile{NoStop}{Balasubramanian:2024rek}%
\bibitem{Marolf:2020xie}%
  \BibitemOpen
  \bibfield{author}{%
  \bibinfo {author} {\bibfnamefont{Donald}\ \bibnamefont{Marolf}}\ and\ \bibinfo {author} {\bibfnamefont{Henry}\ \bibnamefont{Maxfield}},\ }%
  \bibfield{title}{%
  \enquote{\bibinfo {title} {{Transcending the ensemble: baby universes, spacetime wormholes, and the order and disorder of black hole information}},}\ }%
  \bibfield{journal}{%
  \Doi{10.1007/JHEP08(2020)044}{\bibinfo {journal} {JHEP}}\ }%
  \textbf{\bibinfo {volume} {08}},\ \bibinfo {pages} {044} (\bibinfo {year} {2020}),\ \Eprint{http://arxiv.org/abs/2002.08950}{arXiv:2002.08950 [hep-th]}%
  \bibAnnoteFile{NoStop}{Marolf:2020xie}%
\bibitem{Akers:2022qdl}%
  \BibitemOpen
  \bibfield{author}{%
  \bibinfo {author} {\bibfnamefont{Chris}\ \bibnamefont{Akers}}, \bibinfo {author} {\bibfnamefont{Netta}\ \bibnamefont{Engelhardt}}, \bibinfo {author} {\bibfnamefont{Daniel}\ \bibnamefont{Harlow}}, \bibinfo {author} {\bibfnamefont{Geoff}\ \bibnamefont{Penington}},\ and\ \bibinfo {author} {\bibfnamefont{Shreya}\ \bibnamefont{Vardhan}},\ }%
  \bibfield{title}{%
  \enquote{\bibinfo {title} {{The black hole interior from non-isometric codes and complexity}},}\ }%
  \bibfield{journal}{%
  \Doi{10.1007/JHEP06(2024)155}{\bibinfo {journal} {JHEP}}\ }%
  \textbf{\bibinfo {volume} {06}},\ \bibinfo {pages} {155} (\bibinfo {year} {2024}),\ \Eprint{http://arxiv.org/abs/2207.06536}{arXiv:2207.06536 [hep-th]}%
  \bibAnnoteFile{NoStop}{Akers:2022qdl}%
\bibitem{Antonini:2024yif}%
  \BibitemOpen
  \bibfield{author}{%
  \bibinfo {author} {\bibfnamefont{Stefano}\ \bibnamefont{Antonini}}, \bibinfo {author} {\bibfnamefont{Vijay}\ \bibnamefont{Balasubramanian}}, \bibinfo {author} {\bibfnamefont{Ning}\ \bibnamefont{Bao}}, \bibinfo {author} {\bibfnamefont{ChunJun}\ \bibnamefont{Cao}},\ and\ \bibinfo {author} {\bibfnamefont{Wissam}\ \bibnamefont{Chemissany}},\ }%
  \bibfield{title}{%
  \enquote{\bibinfo {title} {{Non-isometry, state dependence and holography}},}\ }%
  \bibfield{journal}{%
  \Doi{10.1007/JHEP02(2025)150}{\bibinfo {journal} {JHEP}}\ }%
  \textbf{\bibinfo {volume} {02}},\ \bibinfo {pages} {150} (\bibinfo {year} {2025}),\ \Eprint{http://arxiv.org/abs/2411.07296}{arXiv:2411.07296 [hep-th]}%
  \bibAnnoteFile{NoStop}{Antonini:2024yif}%
\bibitem{Balasubramanian:2005kk}%
  \BibitemOpen
  \bibfield{author}{%
  \bibinfo {author} {\bibfnamefont{Vijay}\ \bibnamefont{Balasubramanian}}, \bibinfo {author} {\bibfnamefont{Vishnu}\ \bibnamefont{Jejjala}},\ and\ \bibinfo {author} {\bibfnamefont{Joan}\ \bibnamefont{Simon}},\ }%
  \bibfield{title}{%
  \enquote{\bibinfo {title} {{The Library of Babel}},}\ }%
  \bibfield{journal}{%
  \Doi{10.1142/S0218271805007826}{\bibinfo {journal} {Int. J. Mod. Phys. D}}\ }%
  \textbf{\bibinfo {volume} {14}},\ \bibinfo {pages} {2181--2186} (\bibinfo {year} {2005}),\ \Eprint{http://arxiv.org/abs/hep-th/0505123}{arXiv:hep-th/0505123}%
  \bibAnnoteFile{NoStop}{Balasubramanian:2005kk}%
\bibitem{Barbon:2025bbh}%
  \BibitemOpen
  \bibfield{author}{%
  \bibinfo {author} {\bibfnamefont{J.~L.~F.}\ \bibnamefont{Barb{\'o}n}}\ and\ \bibinfo {author} {\bibfnamefont{E.}~\bibnamefont{Velasco-Aja}},\ }%
  \bibfield{title}{%
  \enquote{\bibinfo {title} {{A note on black hole entropy and wormhole instabilities}},}\ }%
  \bibfield{journal}{%
  \Doi{10.1007/JHEP08(2025)103}{\bibinfo {journal} {JHEP}}\ }%
  \textbf{\bibinfo {volume} {08}},\ \bibinfo {pages} {103} (\bibinfo {year} {2025}),\ \Eprint{http://arxiv.org/abs/2502.00769}{arXiv:2502.00769 [hep-th]}%
  \bibAnnoteFile{NoStop}{Barbon:2025bbh}%
\bibitem{Liu:2025xzd}%
  \BibitemOpen
  \bibfield{author}{%
  \bibinfo {author} {\bibfnamefont{Xiaoyi}\ \bibnamefont{Liu}}, \bibinfo {author} {\bibfnamefont{Donald}\ \bibnamefont{Marolf}},\ and\ \bibinfo {author} {\bibfnamefont{Jorge~E.}\ \bibnamefont{Santos}},\ }%
  \bibfield{title}{%
  \enquote{\bibinfo {title} {{Are $S^1\times S^2$ wormholes generic with large sources?}}.}\ }%
   (\bibinfo {month} {10}\ \bibinfo {year} {2025}),\ \Eprint{http://arxiv.org/abs/2510.01325}{arXiv:2510.01325 [hep-th]}%
  \bibAnnoteFile{NoStop}{Liu:2025xzd}%
\bibitem{Israel:1966rt}%
  \BibitemOpen
  \bibfield{author}{%
  \bibinfo {author} {\bibfnamefont{W.}~\bibnamefont{Israel}},\ }%
  \bibfield{title}{%
  \enquote{\bibinfo {title} {{Singular hypersurfaces and thin shells in general relativity}},}\ }%
  \bibfield{journal}{%
  \Doi{10.1007/BF02710419}{\bibinfo {journal} {Nuovo Cim. B}}\ }%
  \textbf{\bibinfo {volume} {44S10}},\ \bibinfo {pages} {1} (\bibinfo {year} {1966})%
  \bibAnnoteFile{NoStop}{Israel:1966rt}%
\bibitem{Emparan:1999pm}%
  \BibitemOpen
  \bibfield{author}{%
  \bibinfo {author} {\bibfnamefont{Roberto}\ \bibnamefont{Emparan}}, \bibinfo {author} {\bibfnamefont{Clifford~V.}\ \bibnamefont{Johnson}},\ and\ \bibinfo {author} {\bibfnamefont{Robert~C.}\ \bibnamefont{Myers}},\ }%
  \bibfield{title}{%
  \enquote{\bibinfo {title} {{Surface terms as counterterms in the AdS / CFT correspondence}},}\ }%
  \bibfield{journal}{%
  \Doi{10.1103/PhysRevD.60.104001}{\bibinfo {journal} {Phys. Rev. D}}\ }%
  \textbf{\bibinfo {volume} {60}},\ \bibinfo {pages} {104001} (\bibinfo {year} {1999}),\ \Eprint{http://arxiv.org/abs/hep-th/9903238}{arXiv:hep-th/9903238}%
  \bibAnnoteFile{NoStop}{Emparan:1999pm}%
\bibitem{Balasubramanian:1999re}%
  \BibitemOpen
  \bibfield{author}{%
  \bibinfo {author} {\bibfnamefont{Vijay}\ \bibnamefont{Balasubramanian}}\ and\ \bibinfo {author} {\bibfnamefont{Per}\ \bibnamefont{Kraus}},\ }%
  \bibfield{title}{%
  \enquote{\bibinfo {title} {{A Stress tensor for Anti-de Sitter gravity}},}\ }%
  \bibfield{journal}{%
  \Doi{10.1007/s002200050764}{\bibinfo {journal} {Commun. Math. Phys.}}\ }%
  \textbf{\bibinfo {volume} {208}},\ \bibinfo {pages} {413--428} (\bibinfo {year} {1999}),\ \Eprint{http://arxiv.org/abs/hep-th/9902121}{arXiv:hep-th/9902121}%
  \bibAnnoteFile{NoStop}{Balasubramanian:1999re}%
\bibitem{Balasubramanian:2025hns}%
  \BibitemOpen
  \bibfield{author}{%
  \bibinfo {author} {\bibfnamefont{Vijay}\ \bibnamefont{Balasubramanian}}\ and\ \bibinfo {author} {\bibfnamefont{Tom}\ \bibnamefont{Yildirim}},\ }%
  \bibfield{title}{%
  \enquote{\bibinfo {title} {{How to Count States in Gravity}},}\ }%
   (\bibinfo {year} {2025}),\ \Eprint{http://arxiv.org/abs/2506.15767}{arXiv:2506.15767 [hep-th]}%
  \bibAnnoteFile{NoStop}{Balasubramanian:2025hns}%
\bibitem{Balasubramanian:2025akx}%
  \BibitemOpen
  \bibfield{author}{%
  \bibinfo {author} {\bibfnamefont{Vijay}\ \bibnamefont{Balasubramanian}}\ and\ \bibinfo {author} {\bibfnamefont{Tom}\ \bibnamefont{Yildirim}},\ }%
  \bibfield{title}{%
  \enquote{\bibinfo {title} {{Observing Spacetime}},}\ }%
   (\bibinfo {month} {9}\ \bibinfo {year} {2025}),\ \Eprint{http://arxiv.org/abs/2509.09763}{arXiv:2509.09763 [hep-th]}%
  \bibAnnoteFile{NoStop}{Balasubramanian:2025akx}%
\bibitem{Cotler:2016fpe}%
  \BibitemOpen
  \bibfield{author}{%
  \bibinfo {author} {\bibfnamefont{Jordan~S.}\ \bibnamefont{Cotler}}, \bibinfo {author} {\bibfnamefont{Guy}\ \bibnamefont{Gur-Ari}}, \bibinfo {author} {\bibfnamefont{Masanori}\ \bibnamefont{Hanada}}, \bibinfo {author} {\bibfnamefont{Joseph}\ \bibnamefont{Polchinski}}, \bibinfo {author} {\bibfnamefont{Phil}\ \bibnamefont{Saad}}, \bibinfo {author} {\bibfnamefont{Stephen~H.}\ \bibnamefont{Shenker}}, \bibinfo {author} {\bibfnamefont{Douglas}\ \bibnamefont{Stanford}}, \bibinfo {author} {\bibfnamefont{Alexandre}\ \bibnamefont{Streicher}},\ and\ \bibinfo {author} {\bibfnamefont{Masaki}\ \bibnamefont{Tezuka}},\ }%
  \bibfield{title}{%
  \enquote{\bibinfo {title} {{Black Holes and Random Matrices}},}\ }%
  \bibfield{journal}{%
  \Doi{10.1007/JHEP05(2017)118}{\bibinfo {journal} {JHEP}}\ }%
  \textbf{\bibinfo {volume} {05}},\ \bibinfo {pages} {118} (\bibinfo {year} {2017}),\ \bibinfo {note} {[Erratum: JHEP 09, 002 (2018)]},\ \Eprint{http://arxiv.org/abs/1611.04650}{arXiv:1611.04650 [hep-th]}%
  \bibAnnoteFile{NoStop}{Cotler:2016fpe}%
\bibitem{Saad:2018bqo}%
  \BibitemOpen
  \bibfield{author}{%
  \bibinfo {author} {\bibfnamefont{Phil}\ \bibnamefont{Saad}}, \bibinfo {author} {\bibfnamefont{Stephen~H.}\ \bibnamefont{Shenker}},\ and\ \bibinfo {author} {\bibfnamefont{Douglas}\ \bibnamefont{Stanford}},\ }%
  \bibfield{title}{%
  \enquote{\bibinfo {title} {{A semiclassical ramp in SYK and in gravity}},}\ }%
   (\bibinfo {month} {6}\ \bibinfo {year} {2018}),\ \Eprint{http://arxiv.org/abs/1806.06840}{arXiv:1806.06840 [hep-th]}%
  \bibAnnoteFile{NoStop}{Saad:2018bqo}%
\bibitem{Saad:2019lba}%
  \BibitemOpen
  \bibfield{author}{%
  \bibinfo {author} {\bibfnamefont{Phil}\ \bibnamefont{Saad}}, \bibinfo {author} {\bibfnamefont{Stephen~H.}\ \bibnamefont{Shenker}},\ and\ \bibinfo {author} {\bibfnamefont{Douglas}\ \bibnamefont{Stanford}},\ }%
  \bibfield{title}{%
  \enquote{\bibinfo {title} {{JT gravity as a matrix integral}},}\ }%
   (\bibinfo {month} {3}\ \bibinfo {year} {2019}),\ \Eprint{http://arxiv.org/abs/1903.11115}{arXiv:1903.11115 [hep-th]}%
  \bibAnnoteFile{NoStop}{Saad:2019lba}%
\bibitem{Penington:2019kki}%
  \BibitemOpen
  \bibfield{author}{%
  \bibinfo {author} {\bibfnamefont{Geoff}\ \bibnamefont{Penington}}, \bibinfo {author} {\bibfnamefont{Stephen~H.}\ \bibnamefont{Shenker}}, \bibinfo {author} {\bibfnamefont{Douglas}\ \bibnamefont{Stanford}},\ and\ \bibinfo {author} {\bibfnamefont{Zhenbin}\ \bibnamefont{Yang}},\ }%
  \bibfield{title}{%
  \enquote{\bibinfo {title} {{Replica wormholes and the black hole interior}},}\ }%
  \bibfield{journal}{%
  \Doi{10.1007/JHEP03(2022)205}{\bibinfo {journal} {JHEP}}\ }%
  \textbf{\bibinfo {volume} {03}},\ \bibinfo {pages} {205} (\bibinfo {year} {2022}),\ \Eprint{http://arxiv.org/abs/1911.11977}{arXiv:1911.11977 [hep-th]}%
  \bibAnnoteFile{NoStop}{Penington:2019kki}%
\bibitem{Chandra:2022bqq}%
  \BibitemOpen
  \bibfield{author}{%
  \bibinfo {author} {\bibfnamefont{Jeevan}\ \bibnamefont{Chandra}}, \bibinfo {author} {\bibfnamefont{Scott}\ \bibnamefont{Collier}}, \bibinfo {author} {\bibfnamefont{Thomas}\ \bibnamefont{Hartman}},\ and\ \bibinfo {author} {\bibfnamefont{Alexander}\ \bibnamefont{Maloney}},\ }%
  \bibfield{title}{%
  \enquote{\bibinfo {title} {{Semiclassical 3D gravity as an average of large-c CFTs}},}\ }%
  \bibfield{journal}{%
  \Doi{10.1007/JHEP12(2022)069}{\bibinfo {journal} {JHEP}}\ }%
  \textbf{\bibinfo {volume} {12}},\ \bibinfo {pages} {069} (\bibinfo {year} {2022}),\ \Eprint{http://arxiv.org/abs/2203.06511}{arXiv:2203.06511 [hep-th]}%
  \bibAnnoteFile{NoStop}{Chandra:2022bqq}%
\bibitem{Sasieta:2022ksu}%
  \BibitemOpen
  \bibfield{author}{%
  \bibinfo {author} {\bibfnamefont{Martin}\ \bibnamefont{Sasieta}},\ }%
  \bibfield{title}{%
  \enquote{\bibinfo {title} {{Wormholes from heavy operator statistics in AdS/CFT}},}\ }%
  \bibfield{journal}{%
  \Doi{10.1007/JHEP03(2023)158}{\bibinfo {journal} {JHEP}}\ }%
  \textbf{\bibinfo {volume} {03}},\ \bibinfo {pages} {158} (\bibinfo {year} {2023}),\ \Eprint{http://arxiv.org/abs/2211.11794}{arXiv:2211.11794 [hep-th]}%
  \bibAnnoteFile{NoStop}{Sasieta:2022ksu}%
\bibitem{deBoer:2023vsm}%
  \BibitemOpen
  \bibfield{author}{%
  \bibinfo {author} {\bibfnamefont{Jan}\ \bibnamefont{de~Boer}}, \bibinfo {author} {\bibfnamefont{Diego}\ \bibnamefont{Liska}}, \bibinfo {author} {\bibfnamefont{Boris}\ \bibnamefont{Post}},\ and\ \bibinfo {author} {\bibfnamefont{Martin}\ \bibnamefont{Sasieta}},\ }%
  \bibfield{title}{%
  \enquote{\bibinfo {title} {{A principle of maximum ignorance for semiclassical gravity}},}\ }%
  \bibfield{journal}{%
  \Doi{10.1007/JHEP02(2024)003}{\bibinfo {journal} {JHEP}}\ }%
  \textbf{\bibinfo {volume} {2024}},\ \bibinfo {pages} {003} (\bibinfo {year} {2024}),\ \Eprint{http://arxiv.org/abs/2311.08132}{arXiv:2311.08132 [hep-th]}%
  \bibAnnoteFile{NoStop}{deBoer:2023vsm}%
\bibitem{Stanford:2020wkf}%
  \BibitemOpen
  \bibfield{author}{%
  \bibinfo {author} {\bibfnamefont{Douglas}\ \bibnamefont{Stanford}},\ }%
  \bibfield{title}{%
  \enquote{\bibinfo {title} {{More quantum noise from wormholes}},}\ }%
   (\bibinfo {month} {8}\ \bibinfo {year} {2020}),\ \Eprint{http://arxiv.org/abs/2008.08570}{arXiv:2008.08570 [hep-th]}%
  \bibAnnoteFile{NoStop}{Stanford:2020wkf}%
\bibitem{Altland:2020ccq}%
  \BibitemOpen
  \bibfield{author}{%
  \bibinfo {author} {\bibfnamefont{Alexander}\ \bibnamefont{Altland}}\ and\ \bibinfo {author} {\bibfnamefont{Julian}\ \bibnamefont{Sonner}},\ }%
  \bibfield{title}{%
  \enquote{\bibinfo {title} {{Late time physics of holographic quantum chaos}},}\ }%
  \bibfield{journal}{%
  \Doi{10.21468/SciPostPhys.11.2.034}{\bibinfo {journal} {SciPost Phys.}}\ }%
  \textbf{\bibinfo {volume} {11}},\ \bibinfo {pages} {034} (\bibinfo {year} {2021}),\ \Eprint{http://arxiv.org/abs/2008.02271}{arXiv:2008.02271 [hep-th]}%
  \bibAnnoteFile{NoStop}{Altland:2020ccq}%
\bibitem{Collier:2019weq}%
  \BibitemOpen
  \bibfield{author}{%
  \bibinfo {author} {\bibfnamefont{Scott}\ \bibnamefont{Collier}}, \bibinfo {author} {\bibfnamefont{Alexander}\ \bibnamefont{Maloney}}, \bibinfo {author} {\bibfnamefont{Henry}\ \bibnamefont{Maxfield}},\ and\ \bibinfo {author} {\bibfnamefont{Ioannis}\ \bibnamefont{Tsiares}},\ }%
  \bibfield{title}{%
  \enquote{\bibinfo {title} {{Universal dynamics of heavy operators in CFT$_{2}$}},}\ }%
  \bibfield{journal}{%
  \Doi{10.1007/JHEP07(2020)074}{\bibinfo {journal} {JHEP}}\ }%
  \textbf{\bibinfo {volume} {07}},\ \bibinfo {pages} {074} (\bibinfo {year} {2020}),\ \Eprint{http://arxiv.org/abs/1912.00222}{arXiv:1912.00222 [hep-th]}%
  \bibAnnoteFile{NoStop}{Collier:2019weq}%
\bibitem{Cotler:2021cqa}%
  \BibitemOpen
  \bibfield{author}{%
  \bibinfo {author} {\bibfnamefont{Jordan}\ \bibnamefont{Cotler}}\ and\ \bibinfo {author} {\bibfnamefont{Kristan}\ \bibnamefont{Jensen}},\ }%
  \bibfield{title}{%
  \enquote{\bibinfo {title} {{Wormholes and black hole microstates in AdS/CFT}},}\ }%
  \bibfield{journal}{%
  \Doi{10.1007/JHEP09(2021)001}{\bibinfo {journal} {JHEP}}\ }%
  \textbf{\bibinfo {volume} {09}},\ \bibinfo {pages} {001} (\bibinfo {year} {2021}),\ \Eprint{http://arxiv.org/abs/2104.00601}{arXiv:2104.00601 [hep-th]}%
  \bibAnnoteFile{NoStop}{Cotler:2021cqa}%
\bibitem{Saad:2019pqd}%
  \BibitemOpen
  \bibfield{author}{%
  \bibinfo {author} {\bibfnamefont{Phil}\ \bibnamefont{Saad}},\ }%
  \bibfield{title}{%
  \enquote{\bibinfo {title} {{Late Time Correlation Functions, Baby Universes, and ETH in JT Gravity}},}\ }%
   (\bibinfo {month} {10}\ \bibinfo {year} {2019}),\ \Eprint{http://arxiv.org/abs/1910.10311}{arXiv:1910.10311 [hep-th]}%
  \bibAnnoteFile{NoStop}{Saad:2019pqd}%
\bibitem{Jafferis:2022uhu}%
  \BibitemOpen
  \bibfield{author}{%
  \bibinfo {author} {\bibfnamefont{Daniel~Louis}\ \bibnamefont{Jafferis}}, \bibinfo {author} {\bibfnamefont{David~K.}\ \bibnamefont{Kolchmeyer}}, \bibinfo {author} {\bibfnamefont{Baur}\ \bibnamefont{Mukhametzhanov}},\ and\ \bibinfo {author} {\bibfnamefont{Julian}\ \bibnamefont{Sonner}},\ }%
  \bibfield{title}{%
  \enquote{\bibinfo {title} {{Matrix Models for Eigenstate Thermalization}},}\ }%
  \bibfield{journal}{%
  \Doi{10.1103/PhysRevX.13.031033}{\bibinfo {journal} {Phys. Rev. X}}\ }%
  \textbf{\bibinfo {volume} {13}},\ \bibinfo {pages} {031033} (\bibinfo {year} {2023}),\ \Eprint{http://arxiv.org/abs/2209.02130}{arXiv:2209.02130 [hep-th]}%
  \bibAnnoteFile{NoStop}{Jafferis:2022uhu}%
\bibitem{Jafferis:2022wez}%
  \BibitemOpen
  \bibfield{author}{%
  \bibinfo {author} {\bibfnamefont{Daniel~Louis}\ \bibnamefont{Jafferis}}, \bibinfo {author} {\bibfnamefont{David~K.}\ \bibnamefont{Kolchmeyer}}, \bibinfo {author} {\bibfnamefont{Baur}\ \bibnamefont{Mukhametzhanov}},\ and\ \bibinfo {author} {\bibfnamefont{Julian}\ \bibnamefont{Sonner}},\ }%
  \bibfield{title}{%
  \enquote{\bibinfo {title} {{Jackiw-Teitelboim gravity with matter, generalized eigenstate thermalization hypothesis, and random matrices}},}\ }%
  \bibfield{journal}{%
  \Doi{10.1103/PhysRevD.108.066015}{\bibinfo {journal} {Phys. Rev. D}}\ }%
  \textbf{\bibinfo {volume} {108}},\ \bibinfo {pages} {066015} (\bibinfo {year} {2023}),\ \Eprint{http://arxiv.org/abs/2209.02131}{arXiv:2209.02131 [hep-th]}%
  \bibAnnoteFile{NoStop}{Jafferis:2022wez}%
\bibitem{Deutsch:1991msp}%
  \BibitemOpen
  \bibfield{author}{%
  \bibinfo {author} {\bibfnamefont{J.~M.}\ \bibnamefont{Deutsch}},\ }%
  \bibfield{title}{%
  \enquote{\bibinfo {title} {{Quantum statistical mechanics in a closed system}},}\ }%
  \bibfield{journal}{%
  \Doi{10.1103/PhysRevA.43.2046}{\bibinfo {journal} {Phys. Rev. A}}\ }%
  \textbf{\bibinfo {volume} {43}},\ \bibinfo {pages} {2046} (\bibinfo {year} {1991})%
  \bibAnnoteFile{NoStop}{Deutsch:1991msp}%
\bibitem{Srednicki:1994mfb}%
  \BibitemOpen
  \bibfield{author}{%
  \bibinfo {author} {\bibfnamefont{Mark}\ \bibnamefont{Srednicki}},\ }%
  \bibfield{title}{%
  \enquote{\bibinfo {title} {{Chaos and Quantum Thermalization}},}\ }%
  \bibfield{journal}{%
  \bibinfo {journal} {Phys. Rev. E}\ }%
  \textbf{\bibinfo {volume} {50}} (\bibinfo {month} {3}\ \bibinfo {year} {1994}),\ \Eprint{http://arxiv.org/abs/cond-mat/9403051}{arXiv:cond-mat/9403051}%
  \bibAnnoteFile{NoStop}{Srednicki:1994mfb}%
\bibitem{Srednicki:1999bhx}%
  \BibitemOpen
  \bibfield{author}{%
  \bibinfo {author} {\bibfnamefont{Mark}\ \bibnamefont{Srednicki}},\ }%
  \bibfield{title}{%
  \enquote{\bibinfo {title} {{The approach to thermal equilibrium in quantized chaotic systems}},}\ }%
  \bibfield{journal}{%
  \Doi{10.1088/0305-4470/32/7/007}{\bibinfo {journal} {J. Phys. A}}\ }%
  \textbf{\bibinfo {volume} {32}},\ \bibinfo {pages} {1163} (\bibinfo {year} {1999})%
  \bibAnnoteFile{NoStop}{Srednicki:1999bhx}%
\bibitem{Pollack:2020gfa}%
  \BibitemOpen
  \bibfield{author}{%
  \bibinfo {author} {\bibfnamefont{Jason}\ \bibnamefont{Pollack}}, \bibinfo {author} {\bibfnamefont{Moshe}\ \bibnamefont{Rozali}}, \bibinfo {author} {\bibfnamefont{James}\ \bibnamefont{Sully}},\ and\ \bibinfo {author} {\bibfnamefont{David}\ \bibnamefont{Wakeham}},\ }%
  \bibfield{title}{%
  \enquote{\bibinfo {title} {{Eigenstate Thermalization and Disorder Averaging in Gravity}},}\ }%
  \bibfield{journal}{%
  \Doi{10.1103/PhysRevLett.125.021601}{\bibinfo {journal} {Phys. Rev. Lett.}}\ }%
  \textbf{\bibinfo {volume} {125}},\ \bibinfo {pages} {021601} (\bibinfo {year} {2020}),\ \Eprint{http://arxiv.org/abs/2002.02971}{arXiv:2002.02971 [hep-th]}%
  \bibAnnoteFile{NoStop}{Pollack:2020gfa}%
\bibitem{Belin:2020jxr}%
  \BibitemOpen
  \bibfield{author}{%
  \bibinfo {author} {\bibfnamefont{Alexandre}\ \bibnamefont{Belin}}, \bibinfo {author} {\bibfnamefont{Jan}\ \bibnamefont{De~Boer}}, \bibinfo {author} {\bibfnamefont{Pranjal}\ \bibnamefont{Nayak}},\ and\ \bibinfo {author} {\bibfnamefont{Julian}\ \bibnamefont{Sonner}},\ }%
  \bibfield{title}{%
  \enquote{\bibinfo {title} {{Charged eigenstate thermalization, Euclidean wormholes and global symmetries in quantum gravity}},}\ }%
  \bibfield{journal}{%
  \Doi{10.21468/SciPostPhys.12.2.059}{\bibinfo {journal} {SciPost Phys.}}\ }%
  \textbf{\bibinfo {volume} {12}},\ \bibinfo {pages} {059} (\bibinfo {year} {2022}),\ \Eprint{http://arxiv.org/abs/2012.07875}{arXiv:2012.07875 [hep-th]}%
  \bibAnnoteFile{NoStop}{Belin:2020jxr}%
\bibitem{deBoer:2024mqg}%
  \BibitemOpen
  \bibfield{author}{%
  \bibinfo {author} {\bibfnamefont{Jan}\ \bibnamefont{de~Boer}}, \bibinfo {author} {\bibfnamefont{Diego}\ \bibnamefont{Liska}},\ and\ \bibinfo {author} {\bibfnamefont{Boris}\ \bibnamefont{Post}},\ }%
  \bibfield{title}{%
  \enquote{\bibinfo {title} {{Multiboundary wormholes and OPE statistics}},}\ }%
  \bibfield{journal}{%
  \Doi{10.1007/JHEP10(2024)207}{\bibinfo {journal} {JHEP}}\ }%
  \textbf{\bibinfo {volume} {10}},\ \bibinfo {pages} {207} (\bibinfo {year} {2024}),\ \Eprint{http://arxiv.org/abs/2405.13111}{arXiv:2405.13111 [hep-th]}%
  \bibAnnoteFile{NoStop}{deBoer:2024mqg}%
\bibitem{Geng:2025efs}%
  \BibitemOpen
  \bibfield{author}{%
  \bibinfo {author} {\bibfnamefont{Hao}\ \bibnamefont{Geng}}, \bibinfo {author} {\bibfnamefont{Ling-Yan}\ \bibnamefont{Hung}},\ and\ \bibinfo {author} {\bibfnamefont{Yikun}\ \bibnamefont{Jiang}},\ }%
  \bibfield{title}{%
  \enquote{\bibinfo {title} {{It from ETH: Multi-interval Entanglement and Replica Wormholes from Large-$c$ BCFT Ensemble}},}\ }%
   (\bibinfo {month} {5}\ \bibinfo {year} {2025}),\ \Eprint{http://arxiv.org/abs/2505.20385}{arXiv:2505.20385 [hep-th]}%
  \bibAnnoteFile{NoStop}{Geng:2025efs}%
\bibitem{Chandra:2023dgq}%
  \BibitemOpen
  \bibfield{author}{%
  \bibinfo {author} {\bibfnamefont{Jeevan}\ \bibnamefont{Chandra}}\ and\ \bibinfo {author} {\bibfnamefont{Thomas}\ \bibnamefont{Hartman}},\ }%
  \bibfield{title}{%
  \enquote{\bibinfo {title} {{Toward random tensor networks and holographic codes in CFT}},}\ }%
  \bibfield{journal}{%
  \Doi{10.1007/JHEP05(2023)109}{\bibinfo {journal} {JHEP}}\ }%
  \textbf{\bibinfo {volume} {05}},\ \bibinfo {pages} {109} (\bibinfo {year} {2023}),\ \Eprint{http://arxiv.org/abs/2302.02446}{arXiv:2302.02446 [hep-th]}%
  \bibAnnoteFile{NoStop}{Chandra:2023dgq}%
\bibitem{Balasubramanian:2006jt}%
  \BibitemOpen
  \bibfield{author}{%
  \bibinfo {author} {\bibfnamefont{Vijay}\ \bibnamefont{Balasubramanian}}, \bibinfo {author} {\bibfnamefont{Bartlomiej}\ \bibnamefont{Czech}}, \bibinfo {author} {\bibfnamefont{Klaus}\ \bibnamefont{Larjo}},\ and\ \bibinfo {author} {\bibfnamefont{Joan}\ \bibnamefont{Simon}},\ }%
  \bibfield{title}{%
  \enquote{\bibinfo {title} {{Integrability versus information loss: A simple example}},}\ }%
  \bibfield{journal}{%
  \Doi{10.1088/1126-6708/2006/11/001}{\bibinfo {journal} {JHEP}}\ }%
  \textbf{\bibinfo {volume} {11}},\ \bibinfo {pages} {001} (\bibinfo {year} {2006}),\ \Eprint{http://arxiv.org/abs/hep-th/0602263}{arXiv:hep-th/0602263}%
  \bibAnnoteFile{NoStop}{Balasubramanian:2006jt}%
\bibitem{Balasubramanian:2025qoz}%
  \BibitemOpen
  \bibfield{author}{%
  \bibinfo {author} {\bibfnamefont{Vijay}\ \bibnamefont{Balasubramanian}}, \bibinfo {author} {\bibfnamefont{Esko}\ \bibnamefont{Keski-Vakkuri}},\ and\ \bibinfo {author} {\bibfnamefont{Nicola}\ \bibnamefont{Pranzini}},\ }%
  \bibfield{title}{%
  \enquote{\bibinfo {title} {{Detector-based measurement-induced state updates in AdS/CFT}},}\ }%
   (\bibinfo {month} {9}\ \bibinfo {year} {2025}),\ \Eprint{http://arxiv.org/abs/2509.13457}{arXiv:2509.13457 [hep-th]}%
  \bibAnnoteFile{NoStop}{Balasubramanian:2025qoz}%
\end{thebibliography}%

\end{document}